\documentclass[12pt,twocolumn,a4paper]{aastex63}
\hypersetup{linkcolor=cyan,citecolor=blue,urlcolor=cyan}
\usepackage{xspace}
\UseRawInputEncoding
\usepackage{amsmath}

\newcommand{\cdig}{C$\rm _{DIG}$\xspace}

\newcommand{\wha}{equivalent width\xspace}

\newcommand{\Q}{ionization parameter\xspace}

\newcommand{\Z}{metallicity\xspace}

\newcommand{\sha}{\rm [S\textsc{II}]/H$\alpha$\xspace}
\newcommand{\hii}{H\,\textsc{II}\xspace}

\newcommand{\shar}{[S\textsc{II}]/H$\alpha$ ratio\xspace}

\usepackage{float}
\usepackage[caption = false]{subfig}
\usepackage{xcolor}

\shorttitle{Measuring the diffuse ionised gas in  galaxies}
\shortauthors{Tomi\v{c}i\'{c} et al.}

\graphicspath{{./}{figures/}}

\begin{document} 

\title{GASP XXXII. Measuring the diffuse ionized gas fraction in ram-pressure stripped galaxies}

\author{Tomi\v{c}i\'{c} Neven}
\affiliation{INAF- Osservatorio astronomico di Padova, Vicolo Osservatorio 5, 35122 Padova, Italy}

\author{Vulcani Benedetta}
\affiliation{INAF- Osservatorio astronomico di Padova, Vicolo Osservatorio 5, 35122 Padova, Italy}

\author{Poggianti  Bianca M.}
\affiliation{INAF- Osservatorio astronomico di Padova, Vicolo Osservatorio 5, 35122 Padova, Italy}

\author{Mingozzi Matilde}
\affiliation{INAF- Osservatorio astronomico di Padova, Vicolo Osservatorio 5, 35122 Padova, Italy}

\author{Werle Ariel}
\affiliation{INAF- Osservatorio astronomico di Padova, Vicolo Osservatorio 5, 35122 Padova, Italy}

\author{Bettoni Daniela}
\affiliation{INAF- Osservatorio astronomico di Padova, Vicolo Osservatorio 5, 35122 Padova, Italy}

\author{Franchetto Andrea}
\affiliation{Dipartimento di Fisica e Astronomia ``Galileo Galilei”, Universit\`a di Padova, vicolo dell’Osservatorio 3, IT-35122, Padova, Italy}
\affiliation{INAF- Osservatorio astronomico di Padova, Vicolo Osservatorio 5, 35122 Padova, Italy}

\author{Gullieuszik  Marco}
\affiliation{INAF- Osservatorio astronomico di Padova, Vicolo Osservatorio 5, 35122 Padova, Italy}

\author{Moretti Alessia}
\affiliation{INAF- Osservatorio astronomico di Padova, Vicolo Osservatorio 5, 35122 Padova, Italy}

\author{Fritz Jacopo}
\affiliation{Instituto de Radioastronom\'ia y Astrof\'isica, UNAM, Campus Morelia, A.P. 3-72, C.P. 58089, Mexico}

\author{Bellhouse Callum}
\affiliation{INAF- Osservatorio astronomico di Padova, Vicolo Osservatorio 5, 35122 Padova, Italy}

\correspondingauthor{Tomi\v{c}i\'{c} Neven}
\email{neven.tomicic@inaf.it}

\begin{abstract}

The diffuse ionized gas (DIG) is an important component of the interstellar medium and it can be affected by many physical processes in galaxies. 
Measuring its distribution and contribution  in emission allows us to properly study both its ionization and star formation in galaxies.
Here, we measure for the first time the DIG emission in 38 gas-stripped galaxies in local clusters drawn from the GAs Stripping Phenomena in galaxies with MUSE survey (GASP). These galaxies are at different stages of stripping. 
We also compare the DIG properties to those of 33 normal galaxies from the same survey.  
To estimate the DIG fraction (C$_{DIG}$) and derive its maps, we combine attenuation corrected H$\alpha$ surface brightness with $\rm [S\textsc{II}]/H\alpha$ line ratio. 
Our results indicate that we cannot use neither a single H$\alpha$ or  $\rm [S\textsc{II}]/H\alpha$ value,  nor a threshold in equivalent width of H$\alpha$ emission line to separate spaxels dominated by  DIG and non-DIG emission. 
Assuming a constant surface brightness of the DIG across galaxies underestimates  C$_{DIG}$.
Contrasting stripped and non-stripped galaxies, we find no clear differences in   C$_{DIG}$.
The DIG emission  contributes  between 20\% and 90\% of the total integrated flux, and does not  correlate with the galactic stellar mass and star-formation rate (SFR). 
The C$_{DIG}$  anti-correlates with the specific SFR, which may indicate an older ($>10^8$ yr) stellar population as ionizing  source of the DIG.
The  DIG fraction shows  anti-correlations with the SFR  surface density, which could be used for a robust estimation of integrated C$_{DIG}$ in galaxies.

\end{abstract}




\section{Introduction} \label{sec:intro}

The interstellar medium (ISM) is an important component of galaxies that determines the galactic evolution and morphology through  star formation (SF),  regulates the  exchange of chemical elements within galaxies, and  affects many physical processes that leave observable signatures in the emitted and absorbed light (\citealt{Schmidt59}, \citealt{Kennicutt98}, \citealt{Kennicutt98b}, \citealt{DraineBook11},   \citealt{Calzetti94}).  
Physical processes outside  and within galaxies, such as interactions, gas stripping, stellar and active galactic nucleus (AGN) feedback, and gravity,  also affect the properties and distribution of various ISM components.
Therefore, observations of the ISM at different wavelengths allows us to trace its different components and track the interplay of different physical mechanisms in galaxies.     

The diffuse ionized gas (DIG), also known as Warm Ionized Medium (WIM) in the Milky Way, is one of the main components of the ISM  (\citealt{Reynolds84}, \citealt{Reynolds92}, \citealt{Walterbos94}, \citealt{Madsen06}, \citealt{Haffner09}, \citealt{Rueff13}, \citealt{Barnes14},  \citealt{Asari20}).  
It is an extended ionized gas between  star forming regions (\hii),  reaching  scale-heights of up to 1-2 kpc in the vertical line of sight from galactic disks, further than a typical size ($\approx50$ pc) of star-forming  associations  (\citealt{Reynolds92}, \citealt{Haffner09}, \citealt{Bocchio16}, \citealt{Tomicic17}). 
Furthermore, the DIG is warmer  (temperatures higher than $\sim10^4$ K) and less dense ($\rho \sim 10^{-1}$ cm$^{-3}$) than the gas
in the \hii regions (\citealt{Collins01}, \citealt{Reynolds01}, \citealt{Haffner09}, \citealt{Barnes14} \citealt{Bruna20}).
The DIG emission has lower surface brightness of the Balmer lines compared to the \hii regions and associations (\citealt{Reynolds84}, \citealt{Reynolds92}, \citealt{Kreckel16}, \citealt{Kumari19}). 
It also shows  higher values of \sha emission line ratios ($\rm [S\textsc{II}]\,\lambda6717,6731/H\alpha>0.2$) compared to the ionized gas in  typical \hii regions ($\rm [S\textsc{II}]\,\lambda6717,6731/H\alpha\approx0.1$), likely due to higher temperatures and lower densities (\citealt{Reynolds84}, \citealt{Madsen06}, \citealt{Blanc09}, \citealt{Kreckel13}).  

The ionizing source of the DIG is not yet conclusively determined, although multiple sources may contribute.  
 The  dominant source may be leaked radiation from the OB stars, whose ionizing photons are able to escape  dusty regions surrounding  \hii regions, thus ionizing gas at larger galactic scale-heights (\citealt{Reynolds92}, \citealt{Minter98}, \citealt{Haffner09}).
The relatively few ionizing photons coming from a large amount of hot, old and low-mass evolved stars (HOLMES) could explain large amounts of the DIG across galaxies  (\citealt{Flores11}, \citealt{Lacerda18}).  
Other ionizing sources are required  to reproduce certain line rations observed in the DIG  (\citealt{Otte02}, \citealt{Hoopes03}).
For example, the DIG  may be heated and ionized by  supernova shocks and turbulence (\citealt{Slavin93}, \citealt{Minter97}) or/and magnetic reconnection (\citealt{Raymond92}).   
According to \citet{Weingartner01}, the electrons from heated dust at larger scale-heights may contribute in ionizing the gas, while \citet{Barnes14} invoked ionization due to  cosmic rays.
Lastly,   \citet{Slavin93} and \citet{Binette09} found in their  models  that the turbulent mixing of the layers of hot and cold gas    could explain the observed line ratios, and higher temperatures. 

In the literature,  the detection and measurements of the DIG fraction (i.e. the fraction of  the gas emission from the DIG to the total gas) is typically done based on one of the following methods: 1) using a threshold in H$\alpha$ surface brightness (using unresolved samples of galaxies or  spatially resolved images), 2) using a threshold in \shar, 3) using the criterion that H$\alpha$  equivalent width is $\rm W_{H\alpha}<3$\,\AA\,  or 4) combining and fitting  the  relation between  the H$\alpha$ surface brightness and $\rm [S\textsc{II}]/H\alpha$ ratio,      (\citealt{Oey07}, \citealt{Blanc09}, \citealt{Kaplan16}, \citealt{Lacerda18},  \citealt{Kreckel16}, \citealt{Zhang17}, \citealt{Poetrodjojo19}, \citealt{Brok20}). 
 Note that the last method assumes the same gas-phase metallicity throughout the galaxy.
This assumption is good enough when a small part of the galaxy is probed, but it might not hold when larger portions of galaxies are investigated. In this case, metallicity variations within  the galaxy, which affect the \sha ratio (\citealt{Blanc15}), should be taken into account. 

Observations show that the DIG  emission in galaxies accounts for  between 20\% and 80\% of the H$\alpha$ emission and covers a large fraction of the galactic area (\citealt{Hoopes99}, \citealt{Oey07}, \citealt{Sanders17}, \citealt{Kreckel16}, \citealt{Bruna20}).
The importance of measuring the DIG distribution and brightness in galaxies  lays in the fact that it may hinder measurements of various physical properties of galaxies. 
For example, since the DIG's source of ionization  may not come from the star-forming regions,  values of  star-formation rates (SFRs) may be overestimated if the DIG contribution, coming from sources other than star formation, is not removed. 
Furthermore, different relative distributions of dust, young stars and extended ionized gas  may lead to a miscalculation of  the attenuation values (A$_V$) of  star-forming regions (\citealt{Calzetti94}, \citealt{Tomicic17}, \citealt{Tomicic19}).

The observations of the DIG allow us to trace  the  physical processes  affecting  galaxies and their ISM, study stellar feedback affecting the surrounding ISM (\citealt{Barnes14}, \citealt{Vandenbroucke19}), and the effects of  collision and turbulent mixing of gas layers with different properties  (\citealt{Slavin93}, \citealt{Binette09}, \citealt{Fumagalli14}, \citealt{Poggianti19b}).
An interesting population of galaxies that might have their DIG affected by the environment includes galaxies that enter galaxy clusters, passing through a  intracluster medium. 
Their gas is being stripped from the disk,  due to the ram-pressure (RP). 
This stripped gas, also seen as ionized gas tails outside the galaxy disks,  collapses and forms new stars.    
Observations of this kind of  hydrodynamical interaction  offer an opportunity to probe the physics of the ISM in different environments and study how it affects  galaxy evolution. 
This includes investigating the DIG distribution, estimating the fraction of the surface brightness coming from the DIG,  and the sources of ionization of the DIG.   

The GASP project (GAs Stripping Phenomena in galaxies with MUSE; \citealt{Poggianti17})  has optical spectral observations of 114 galaxies, of which some are at different levels of the ram-pressure stripping process.
\cite{Poggianti19b} found that approximately 50\% of the $\rm H\alpha$ emission 
in the debris tails of a subsample of GASP  galaxies  is diffuse, i.e. is not related to identifiable star-forming clumps.
In their work, the star-forming clumps  were determined based on the Balmer $\rm H\alpha$ line images  (\citealt{Poggianti17}). 
These star-forming knots were used for evaluation of various properties of the  galaxies and the \hii regions, such as SFRs, gas-phase metallicity variations, source of ionisation etc.  (\citealt{George18},  \citealt{Vulcani19},  \citealt{Poggianti19}, \citealt{Vulcani20b}).

The purpose of this paper is to measure the distribution and fraction of the DIG across the GASP galaxies. 
This paper follows a number of papers investigating statistical properties of the GASP sample (e.g. \citealt{Jaffe18}, \citealt{Vulcani18}, \citealt{Poggianti19}, \citealt{Poggianti17b},  \citealt{Vulcani19}, \citealt{Franchetto20}, \citealt{Gullieuszik20},  \citealt{Vulcani20b}).
We  present for the first time a comparison of the DIG fractions between galaxies  that are being stripped and those which are not. 
The method that we use for estimating the DIG fraction  combines both the H$\alpha$ surface brightness and the \sha ratio,  also taking into account the metallicity variation within galaxies.
We  compare our results for the DIG fraction with the  fractions derived based on the method using only the H$\alpha$ surface brightness, and look at how the \sha line ratio  and equivalent width of H$\alpha$ line vary with the DIG fractions.

This paper is structured as follows. 
We describe the GASP project and the galaxy sample in Sec. \ref{sec:Data}. In the same section, we explain observations, data reduction and spectral analysis. 
Furthermore, we explain how we  derived the emission line maps, gas-phase metallicities and equivalent width. 
In Sec. \ref{sec:estimating Cdig}, we explain the method for estimating the fraction of the DIG.
The results are presented in Sec. \ref{sec:Results}, and the implications of using various methods for estimating the DIG fraction are discussed in Sec. \ref{sec: Discussion}.
Here, we will also show a comparison of the DIG fractions between the stripped and non-stripped galaxies.  
The conclusions and summary are written in Sec. \ref{sec: Summary}. 

In this paper we adopted standard cosmological constants of H$_0=70$ km$\rm \,s^{-1}Mpc^{-1}$,  $\rm \Omega_M=0.3$, $\Omega_\Lambda=0.7$, and the initial mass function (IMF) from \citet{Chabrier03}.


\section{Data} \label{sec:Data}

\begin{figure*}
\centering
 \includegraphics[width=0.95\textwidth]{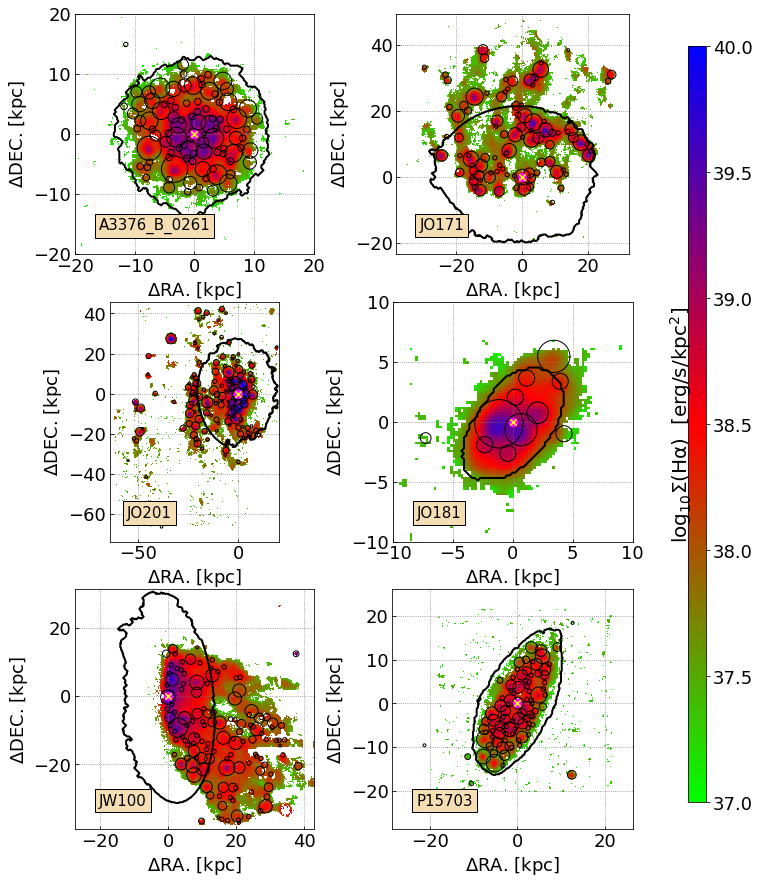}
   \caption{ Maps of the observed surface brightness of H$\alpha$ emission  ($\Sigma H\alpha$) of some galaxies in the GASP survey. The following galaxies are examples of a control sample galaxy ($\rm A3376\_B\_0261$), gas stripped galaxies (ring galaxy JO171, face-on JO201, low gas-phase metallicity JO181, and JW100), and a field galaxy (P15703). In each frame, we indicate the stellar disk (thick, black contour; \citealt{Gullieuszik20}), galactic center (yellow cross on magenta circle),  and the $\rm H\alpha$ clumps (black circles; \citealt{Poggianti17}). }
    \label{fig:Fig_Ha}
\end{figure*}


\subsection{Galaxy sample}\label{subsec:Data, gasp}

For this work, we make use of the observations obtained in the context of the multi-wavelength  GASP\footnote{\url{https://web.oapd.inaf.it/gasp/index.html}}  project (\citealt{Poggianti17}).  The  survey targeted 114 late type galaxies in the redshift regime $0.04<z<0.1$, with galaxy stellar masses in the range $10^9< M_\ast/M_\odot<10^{11.5}$ and  located in different environments (galaxy clusters, groups, filaments and isolated). Galaxies in clusters were selected from the  WINGS (\citealt{Fasano06}) and OMEGAWINGS (\citealt{Gullieuszik15}) surveys,  galaxies in the less dense environments are  from the PM2GC catalog (\citealt{Calvi11}). 
GASP includes both galaxies selected as stripping candidates and undisturbed galaxies, plus a few passive galaxies. 

In this work, we  consider only galaxies showing emission lines in their spectra, and exclude interacting galaxies. 
 We  consider separately a stripping sample and a reference sample (i.e. control sample).
 The former includes galaxies with signs of mild, moderate, and extreme stripping, as well as truncated disks, for a total of 38 galaxies.
 We refer to Table 2 in \citet{Vulcani18} for the list of the objects, along with redshifts, coordinates, integrated stellar masses and star formation rates.
 The galaxies with  tails of length comparable to their stellar disks will be labeled as jellyfish galaxies.
 Of  the sample described in \citet{Vulcani18}, we add JO93 for which a careful inspection of its Halpha map indicates an initial phase of stripping. In addition we exclude JO149 and JO95 from this work, since we are not able to measure their effective galactocentric radii and orientations (\citealt{Franchetto20}). 
The stripping sample is made of 38 galaxies.

The control sample includes cluster+field galaxies that are undisturbed and do not show any clear  sign of environmental effects (ram pressure stripping, tidal interaction, mergers, gas accretion, or other interactions) on their spatially resolved star formation distribution, for a total of 33  galaxies, 17 of which are cluster members and 16 field galaxies.
Table 2 of \citet{Vulcani18} presents the galaxies included in the control sample.
From this list, we exclude P19482 because  a subsequent analysis has revealed that the galaxy is most likely undergoing cosmic web enhancement \citep{Vulcani19a}. 
Overall, in what follows we will analyze 71 galaxies.


\subsection{Observations and emission line maps}\label{subsec:Obs, Data, line maps}

A detailed description of the  GASP observations and data reduction can be found in \citet{Poggianti17}. 

The GASP project used  integral field unit (IFU) data, observed with the MUSE instrument (Multi Unit Spectroscopic Explorer),   that provide  spatially resolved, spectroscopic information of  galaxies.  
The spectra in each spaxel of the observed data was corrected for the effect of foreground extinction of light, which is caused by the  Milky Way dust. 
 We used  $E_{B-V}$ values for each galaxy that is measured in the corresponding line of sight (LOS) on sky  by  \citealt{Schlafly11},   and the extinction curve measured by \citet{Cardelli89} assuming $\rm R_V = 3.1$.
 
To account for seeing effects, the data were smoothed and convolved  in the spatial dimension using a $5\times5$ pixels kernel,  which corresponds to $\approx$1 arcsec or 0.7-1.3 kpc depending on the galaxy redshift.   
These convolved cubes were analyzed with the spectrophotometric
code SINOPSIS (\citealt{Fritz17}) that is fitting the stellar spectra with a  combination of single stellar population (SSP) templates  of different ages.
After subtracting the stellar continuum, the emission line cubes are fitted using KUBEVIZ (\citealt{Fossati16}). 
The emission lines taken into account in this work are: $\rm H\beta$, $\rm [O\,\textsc{iii}]\lambda5007$, $\rm [O\,\textsc{i}]\lambda6300$, $\rm H\alpha$, $\rm [N\,\textsc{ii}]\lambda6584$, and $\rm [S\textsc{II}]\lambda6713,6731$. 
In the following sections, we refer to the  $\rm [S\textsc{II}]\,\lambda6717,6731$ doublet  as  $\rm [S\textsc{II}]$. 

The final emission line maps  of H$\alpha$ for a subset of the sample are shown in Fig. \ref{fig:Fig_Ha}. 
The \wha is calculated as the ratio between the H$\alpha$ surface brightness and the stellar continuum near the line, as measured by KUBEVIZ. 

In this work, we will individually present only 6 galaxies to show the most representative examples with different characteristics (inclination, morphology, gas-phase metallicities, stellar masses, ...). 
To describe our analysis we focus on 6 representative galaxies i.e. : from the control sample, $\rm A3376\_B\_0261$  as the best example of a face-on galaxy, and P15703 as an example of an inclined galaxy.
The four gas stripped galaxies are: JO171 - a face-on  ring galaxy,  JO201 - a galaxy with low inclination, JO181 - a galaxy with a low gas-phase metallicity, and  JW100 - a galaxy with a high gas-phase metallicity. 
The maps for the whole sample are shown in appendix.


\subsection{Attenuation, Gas-phase metallicity and BPT diagrams}\label{subsec:Data, PYQZ}

We corrected the emission line maps for the internal dust attenuation (A$_V$) in galaxies using the extinction curve of \citet{Cardelli89}, assuming a foreground-screen dust/gas distribution (\citealt{Calzetti94}, \citealt{Kreckel13}, \citealt{Tomicic17}) and R$\rm _V=3.1$. 
We assumed the intrinsic Balmer line ratio $\rm H\alpha/H\beta=2.86$ of the star forming regions that corresponds to an ionized gas temperature of
$\rm T\approx10^4$ K  and case B recombination (\citealt{Osterbrock92}).  

The maps of $\rm H\alpha$ surface brightness corrected for attenuation, labeled as $\rm \Sigma H\alpha,corr$,  were used in Sec. \ref{sec:estimating Cdig} for calculating the fraction of the diffuse ionized gas. 
We applied a cut in signal-to-noise $\rm S/N>=$4 for the $\rm [S\textsc{II}]$ doublet and H$\beta$ lines, and  $\rm S/N>=$8 for $\rm H\alpha,corr$, to the spaxels used in measuring the surface brightness fraction of the DIG (Sec. \ref{sec:estimating Cdig}). In this way, we discarded the spaxels with high uncertainty but kept the DIG dominated spaxels with potentially weak emission of the weaker lines. 
Spaxels with negative attenuation values were also removed from the estimation of the DIG emission fraction. 
To convert values of $\rm \Sigma H\alpha,corr$ into SFR surface density, we used the SFR prescription defined by \citet{Kennicutt98}, $\rm SFR(M\_{\odot}/yr)=4.6\,\times\,10^{-42}\,L_{H\alpha}(erg/s)$ (\citealt{Poggianti17}). 

The gas-phase metallicity was derived from the emission line ratios using the PYQZ code (\citealt{Dopita13}, \citealt{Vogt15}).  
We used a modified version of PYQZ v0.8.2, the model grid projected on the line-ratio plane $\rm [O\,\textsc{iii}]\lambda5007$/$[S\textsc{II}]$ vs.  $\rm [N\,\textsc{ii}]\lambda6583$/$[S\textsc{II}]$, and the solar metallicity of $\rm 12+log(O/H)=8.69$.
Details of estimating the gas-phase metallicity are described in detail by \citet{Franchetto20}.

 We used the diagnostic Baldwin, Phillips \& Telervich diagram (BPT; \citealt{Baldwin81})   to determine the excitation mechanism of the bright nebular lines, such as emission dominated by  star-formation, composite,  AGN, or LINER\footnote{Low Ionisation Nuclear Emission Regions (\citealt{Heckman80}).}/LIER\footnote{Low Ionisation Emission Regions (\citealt{Belfiore16}).}. 
 The   BPT diagram is  based on  $\rm [O\,\textsc{i}]\lambda6300$   line, and we removed spaxels dominated by the AGN emission during the process of estimating the fraction of the DIG in individual galaxies (in Sec. \ref{sec:estimating Cdig}).\footnote{We did not use BPT diagrams based on  $\rm [N\,\textsc{ii}]$ and $\rm [S\,\textsc{ii}]$ lines because those lines may  be  affected by metallicity variations.}
 Note that we will also use the  spaxels that show composite or LINER/LIER source of ionisation, because those sources  may  potentially indicate an additional origin of the DIG in the tails of the galaxies.
Examples of the BPT diagrams of the GASP galaxies are presented in \citet{Poggianti19}.


\subsection{H$\alpha$ clumps}\label{subsec: clumps}

Galaxies are typically characterized by areas with bright $\rm H\alpha$ emission. A number of GASP papers have already characterized the general properties of these `$\rm H\alpha$ knots' or `$\rm H\alpha$ clumps' (stellar mass, specific SFRs, kinematics, source of ionisation, metallicities, etc., \citealt{Poggianti17}, \citealt{Poggianti19}, \citealt{Vulcani19},  \citealt{Vulcani20b}).
In these works, the $\rm H\alpha$ clumps in the H$\alpha$ emission-only, dust-corrected surface brightness maps were identified by convolving those maps with a Laplacian filter  (using IRAF-laplace) and median filtered (using IRAF-median tools).
Minima in the filtered images were designated as centers of the $\rm H\alpha$ clumps. 
 The radii of the  clumps were estimated in an iterative way through a recursive analysis of three  consecutive circular shells with thickness of 1 pixel around each knot center.
The iteration stopped at the radius at which: 1) there is no more decrease in  surface brightness, or 2) the surface brightness reached the value previously set for the background emission, or 3) when the circular shell reached another peak or the edge of the image.
The procedure of locating and measuring the size of the  clumps is described in detail in  \citet{Poggianti17}.  


\subsection{Galactic disk}
\label{subsec:Data, disk   clumps}

Estimation  of the boundary of stellar disks of galaxies is described in detail by \citet{Poggianti17} and  by \citet{Gullieuszik20} (Sec. 3.1 in their paper).  
This estimation utilises an isophote in continuum map, that is 1$\sigma$ above the average sky  background noise.
We define the spaxels outside the stellar disks as part of  tails. 
The resulting stellar disk boundaries are shown in Fig. \ref{fig:Fig_Ha}, \ref{fig:Cdig_clumps_maps} and \ref{fig:Cdig_clumps_maps_2} as thick contours. 
The galactic centers were designated to the centroids of the brightest central region in the continuum maps (\citealt{Gullieuszik20}). 
We use only disk's data when we compare stripped galaxies with the control sample, as by definition the control sample galaxies do not have tails.

The orientation (the position angle and inclination) and galactocentric radii  were estimated by \citet{Franchetto20} from the I-band images (Sec. 3.1 in their paper). We use these values in Sec. \ref{subsec:estimating Cdig step by step} where we estimated radial decrease in gas-phase metallicities, and in  Sec.  \ref{subsec:Discussion, Wha}, where we correct the data  from the disk for  inclination effects.

\begin{figure*}
\centering
 \includegraphics[width=0.98\textwidth]{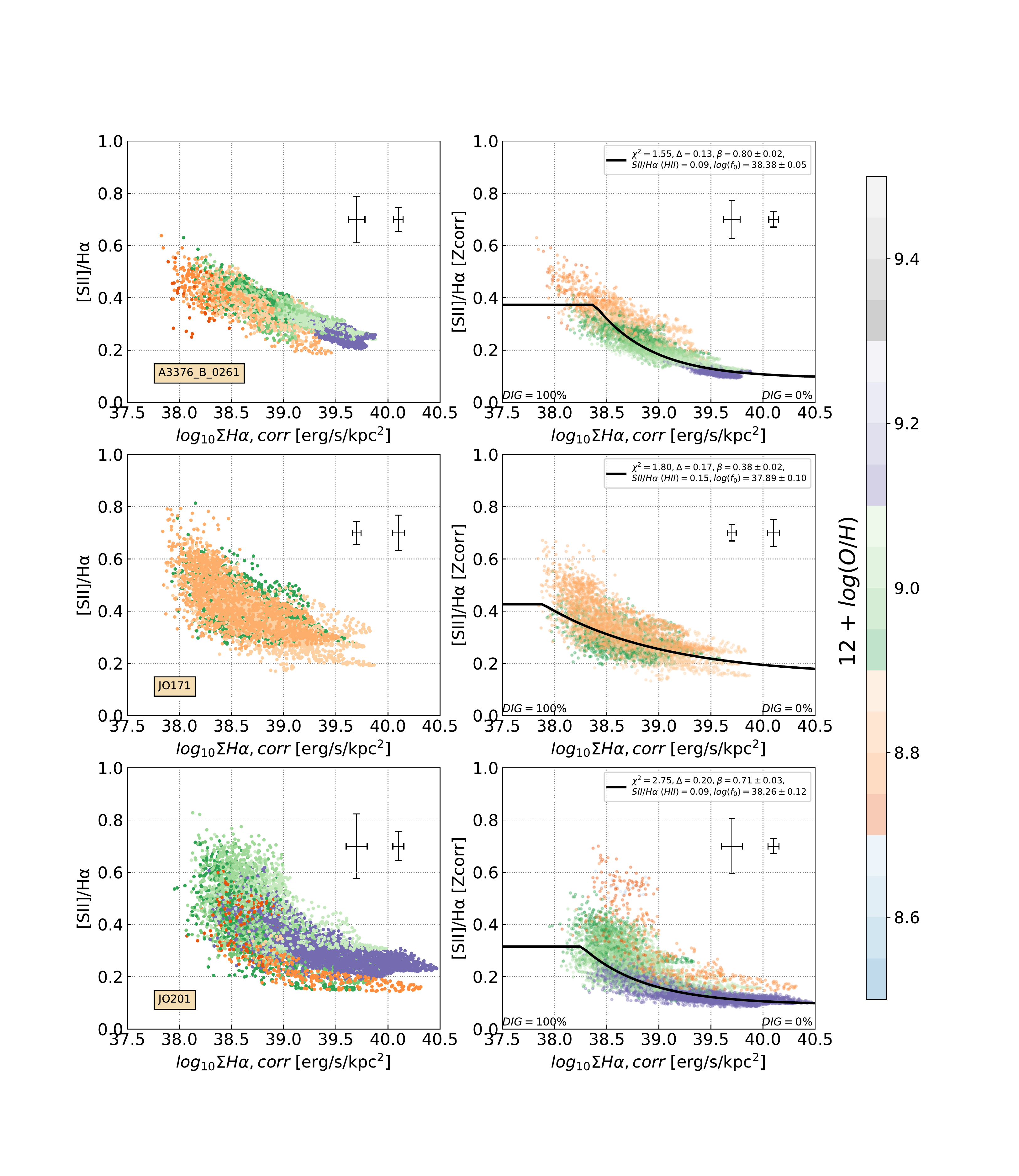}
   \caption{  \textit{Left-} The $\rm [S\textsc{II}]/H\alpha$ vs. $\rm \Sigma H\alpha,corr$ where we did not correct \sha values for metallicity variation.  \textit{Right-} The $\rm [S\textsc{II}]/H\alpha|_{Zcorr}$ vs. $\rm \Sigma H\alpha,corr$, where we did correct \sha values for metalicities that  change with galactocentric radius. The values in all panels were color-coded by   metallicity values that are median values of metallicities within bins of galactocentric radii. On  right panels, we show  fits on the data with the black lines.  We present the median of uncertainties of all data with  error-bar on the right, while  the median of uncertainties of 5\%  spaxels with lowest $\rm \Sigma H\alpha,corr$ with   error-bars on the left.    Also, in the upper right corner, we write $\chi^2$ of the fit, 3$\sigma$  scatter in \sha  of the data from the fitted line ($\Delta$), estimated $\beta$ and $\rm f_0$ values (and their 3$\sigma$ uncertainties), and the  \sha of \hii dominated spaxels. Values of $\rm f_0$ are in the units of surface brightness. Galaxies presented here are $A3376\_B\_0261$, JO171, and JO201.    For details, see the text in Sec \ref{sec:estimating Cdig}.    }
    \label{fig:Fig_SHa_vs_Hac_p1}
\end{figure*}

\begin{figure*}
\centering
 \includegraphics[width=1\textwidth]{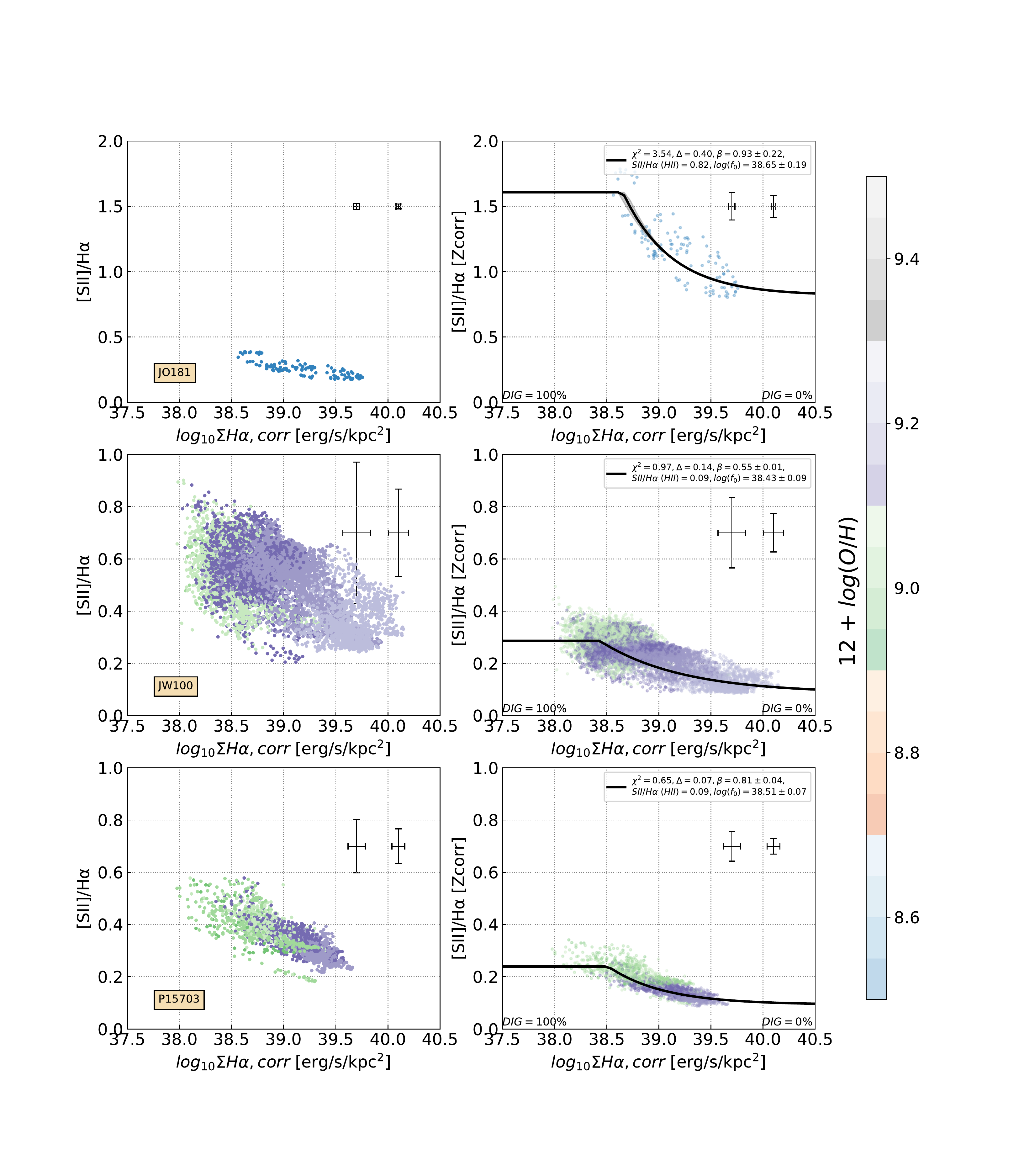}
   \caption{ Same as Fig. \ref{fig:Fig_SHa_vs_Hac_p1}, but for galaxies JO181, JW100 and P15703. }
    \label{fig:Fig_SHa_vs_Hac_p2}
\end{figure*}


\section{Estimating the diffuse ionized gas fraction}
\label{sec:estimating Cdig}

Estimating the  distribution and the  fraction of the diffuse ionized gas in each galaxy is hindered by the fact that generally the observed H$\alpha$ emission in the line of sight  is composed of both the emission coming from regions of dense gas\footnote{ Here we label `dense gas'  what in the literature is labeled as gas from the HII regions. We choose this definition  because we cannot resolve single HII regions due to GASP spatial resolution ($\approx 1$ kpc). } (non DIG)  and from the DIG. 
The fraction of the total H$\alpha$ emission  from the DIG and the dense gas  is labeled as $\rm C_{DIG}$ and $\rm C_{dense}$, respectively, with a mutual relation $\rm C_{DIG}$ = 1 - $\rm C_{dense}$. 
Therefore, following \citet{Blanc09} and \citet{Kaplan16}, we can empirically estimate $\rm C_{DIG}$ across a galaxy, and the total observed H$\alpha$ surface brightness along the LOS ($\rm \Sigma H\alpha,obs$)  relate to the observed H$\alpha$ surface brightness from the DIG ($\rm \Sigma H\alpha,obs |_{DIG}$) as in the following equation:

\begin{equation}
 \rm \Sigma H\alpha,obs =  C_{dense}\cdot \Sigma H\alpha,obs + C_{DIG}\cdot \Sigma H\alpha,obs \;\;\;,
\label{eq:Eq01}
\end{equation}

\begin{equation}
\rm  \Sigma H\alpha,obs |_{DIG} =  C_{DIG}\cdot \Sigma H\alpha,obs \;\;\;.
\label{eq:Eq02}
\end{equation}

\noindent
\citet{Blanc09} and \citet{Kaplan16} compared  $\rm \Sigma(H\alpha,corr)$ and \sha line ratios   and assumed a relation between the \sha ratios and $\rm C_{DIG}$  that depends on the gas phase metallicity:

\begin{equation}
\rm \frac{[S\textsc{II}]}{H\alpha}\Bigg|_{obs} = Z'\left(C_{dense}\cdot \frac{[S\textsc{II}]}{H\alpha}\Bigg|_{dense,\,Zcorr} + C_{DIG}\cdot \frac{[S\textsc{II}]}{H\alpha}\Bigg|_{DIG,\,Zcorr}  \right)  
\label{eq:Eq03}
\end{equation}

\noindent
where $\rm [S\textsc{II}]/H\alpha|_{obs}$ is the observed line ratio in LOS, and $\rm [S\textsc{II}]/H\alpha|_{dense,\,Zcorr}$ and $\rm [S\textsc{II}]/H\alpha|_{DIG,\,Zcorr}$ are empirically determined for  dense gas and DIG dominated spaxels, with the local ISM gas-phase metallicity.
In the Milky Way, the \hii regions (regions of dense gas) show $\rm [S\textsc{II}\,\lambda6717]/H\alpha|_{\hii}=0.11$ with a small scatter (standard deviation  $ \approx0.03$), while the DIG regions show $\rm [S\textsc{II}\,\lambda6717]/H\alpha|_{DIG}\approx0.34$ with a large scatter  (standard deviation $\approx0.13$; \citealt{Madsen06}). 
The Z$'$ value is the ratio of the metallicity of the observed galaxies and the Milky Way ($\rm Z'= Z_{gal}/Z_{MW}$). 
In this case, the Milky Way metallicity values are equivalent to the values of the ISM around the Sun (\citealt{Grevesse96}).

Note that \citet{Blanc09} and \citet{Kaplan16} used one value of Z$'$ per galaxy.
This assumption is good enough when observations cover a relatively small area within  galaxies, but would fail for observations  covering the entire galaxies, including the debris tails outside the stellar disks, as in the case of  GASP. 
Indeed, the strong metallicity variations observed across the galaxy disks (\citealt{Franchetto20}, \citealt{Franchetto20b}) may drastically affect \sha line ratios and thus hinder a proper estimation of the $\rm C_{DIG}$.


\subsection{Step-by-step estimation of $\rm C_{DIG}$}
\label{subsec:estimating Cdig step by step}

In this paper,  we introduce a technique that accounts  for  metallicity variations within   galaxies, unlike the technique previously used by  \citet{Blanc09} and \citet{Kaplan16}. 
Our technique considers the metallicity at each spaxel,  including those in the tails: given the broad range in positions and physical conditions, metallicity might assume a large range. 
We describe in the following subsection the technique for measuring  $\rm C_{DIG}$.

Even though we have gas-phase metallicity values for each spaxel separately, to reduce the noise, we  used  metallicities estimated in the following way. 
We divided the spaxels into annuli of different deprojected galactocentric radii, and estimated a median value of the spaxel metallicities (measured by \citealt{Franchetto20})  for each individual radial annulus. 
Then we assigned to all the spaxels in a given annulus the same measured median metallicity. 
This metallicity value is used as radial metallicity ($Z_{gal}$) for spaxels in their corresponding radial bins, as shown by the colored points in Fig. \ref{fig:Fig_SHa_vs_Hac_p1} and \ref{fig:Fig_SHa_vs_Hac_p2}.

In the first step, we divide the $\rm [S\textsc{II}]/H\alpha$ values by the corresponding    $\rm Z'$ ratio  ($\rm Z'= Z_{gal}/Z_{MW}$). 
We label these new values as $\rm [S\textsc{II}]/H\alpha|_{Zcorr}$. 
In the left panels of  Fig. \ref{fig:Fig_SHa_vs_Hac_p1} and \ref{fig:Fig_SHa_vs_Hac_p2}, we show  $\rm [S\textsc{II}]/H\alpha$  vs. $\rm \Sigma H\alpha,corr$ for 6 selected galaxies. 
In the right panels, we present $\rm [S\textsc{II}]/H\alpha|_{Zcorr}$ as a function of H$\alpha,corr$, and color-coded the spaxels  by radial metallicity. 
The relation \ref{eq:Eq03} then becomes:

\begin{equation}
\rm \frac{[S\textsc{II}]}{H\alpha}\Bigg|_{Zcorr} = C_{dense}\cdot \frac{[S\textsc{II}]}{H\alpha}\Bigg|_{dense,\,Zcorr} + C_{DIG}\cdot \frac{[S\textsc{II}]}{H\alpha}\Bigg|_{DIG,\,Zcorr}  \;\;\;.
\label{eq:Eq03b}
\end{equation}

 \noindent
 As seen on the figure, the scatter of the data after the metallicity correction becomes smaller, thus improving estimation of the DIG fractions.   
 
In the second step, we  empirically estimate  $\rm [S\textsc{II}]/H\alpha|_{DIG,\,Zcorr}$ and $\rm [S\textsc{II}]/H\alpha|_{dense,\,Zcorr}$ as the median value of 5\% spaxels with lowest  and highest value in $\rm \Sigma H\alpha,corr$, assuming that these extreme regimes are dominated by the DIG and dense gas  respectively.
Following the method from \citet{Kaplan16}, we compute a first empirical guess for $\rm C_{DIG}$ relation, which is directly derived from Eq. \ref{eq:Eq03b}, as:

\begin{equation}
\rm C_{DIG} = \frac{[S\textsc{II}]/H\alpha|_{dense,\,Zcorr}-[S\textsc{II}]/H\alpha|_{Zcorr}}{[S\textsc{II}]/H\alpha|_{dense,\,Zcorr}-[S\textsc{II}]/H\alpha|_{DIG,Zcorr}} \;\;\;.
\label{eq:Eq04}
\end{equation}

\noindent
Finally, we fit these estimated $\rm C_{DIG}$ and $\rm \Sigma H\alpha,corr$,  following an assumed relation between $\rm \Sigma H\alpha,corr$ and $\rm C_{DIG}$ (following \citealt{Blanc09} and \citealt{Kaplan16}) as:

\begin{equation}
C_{DIG} = \Big( \frac{f_0}{\Sigma H\alpha,corr} \Big)^{\beta} \;\;\;.
\label{eq:Eq05} 
 \end{equation}

\noindent
This relation  is valid only for $\rm \Sigma H\alpha,corr > f_0$.
Here, $\rm f_0$ indicates the $\rm \Sigma H\alpha,corr$ value for spaxels with DIG-only emission ($\rm C_{DIG}=1$).
In the case that $\beta$ is equal to 1, it would mean that   the DIG surface brightness is constant across the galaxy ($\rm C_{DIG=\, f_0}$),  which follows after combining Eq. \ref{eq:Eq02} and \ref{eq:Eq05}. 
If the DIG is affected by the intensity of the dense gas, $\beta$ would be lower than 1.

The final fit on  $\rm [S\textsc{II}]/H\alpha|_{Zcorr}$ values that correlate with $\rm C_{DIG}$ following  Eq. \ref{eq:Eq03} and \ref{eq:Eq04} is shown as a thick line in the right panels of Fig. \ref{fig:Fig_SHa_vs_Hac_p1} and \ref{fig:Fig_SHa_vs_Hac_p2}. 
We used the MPFIT\footnote{http://purl.com/net/mpfit} (\citealt{MPFIT09}) model in python code for fitting the data.  
Here, the horizontal part of the line ($\rm \Sigma H\alpha,corr < f_0$) assumes the values in $\rm \Sigma H\alpha,corr$ with $\rm C_{DIG}=1$, while the part toward higher $\rm \Sigma H\alpha,corr$ corresponds to $\rm C_{DIG}=0$. 
In the upper right corners of the right panels, we write  $\chi^2$ of the fit, 3$\sigma$  scatter in \sha  of the data from the fitted line ($\Delta$), estimated $\beta$ and $\rm f_0$ values, and the  \sha of dense gas dominated spaxels.

For each individual galaxy, we derive single values of $\rm f_0$ and $\beta$.
Then, following Eq. \ref{eq:Eq05}, we use  $\rm f_0$ and $\beta$ values   on the $\rm \Sigma H\alpha,corr$ map to derive  $\rm C_{DIG}$ in each spaxel.  We show and discuss these maps in Sec. \ref{subsec:Results, Cdig map}. 
Uncertainties in $\rm C_{DIG}$ for individual spaxels are estimated combining the uncertainties from different sources. 
These include errors on variables ($\rm f_0$ and $\beta$, as an output of the fit) and uncertainties of $\rm \Sigma H\alpha,corr$. 
An additional uncertainty is a difference between  the $\rm C_{DIG}$ estimated using spaxel-by-spaxel metallicity and $\rm C_{DIG}$ estimated using metallicity that changes with galactocentric radius.

\begin{figure}
\centering
 \includegraphics[width=0.45\textwidth]{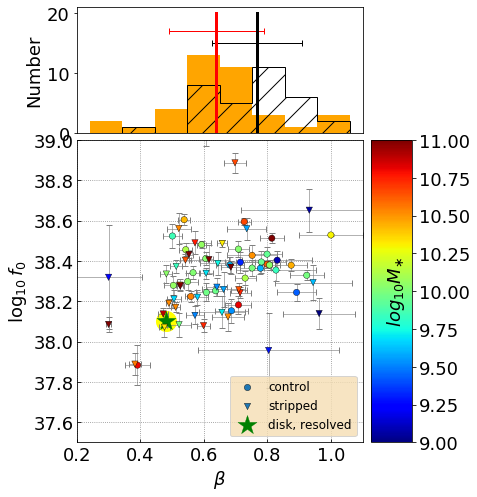}
   \caption{ $\beta$ and $\rm f_0$ values for galaxies of the  control sample (circles) and stripped galaxies (triangles), color-coded by stellar mass.
    The $\rm 3\sigma$ uncertainties are presented with error bars. The green star shows the result of the fit from the resolved data from the disks only (from Fig. \ref{fig:Fig_Resolved_Cdig}).  The histograms show the distribution of  $\beta$ values for the control sample (white with lines) and for stripped galaxies (orange). The median  $\beta$  value (and corresponding $1 \,\sigma$ of distributions) for control (stripped) galaxies is presented with a black (red) line. }
    \label{fig:beta_f0_galaxies}
\end{figure}

 \begin{table*}
\centering
\caption{ Galaxies from the control sample, and their disk $\rm \Sigma SFR$,  $\rm f_0$  and  $\beta$ values (and their  $\rm 3\sigma$ uncertainties), $\rm C_{DIG,\,disk}$ (uncertainty estimated as a mean of spaxel-by-spaxel $\rm C_{DIG}$ uncertainties), and the spatial fractions of spaxels that have $\rm C_{DIG}>0.3$ within the disk (uncertainties calculated assuming a bimodal distribution).  } 
\begin{tabular}{ccccccc}
\\
Name   &   $\rm \Sigma SFR_{disk}$ & $\beta$  & $\rm log_{10}f_0$ & C$\rm _{DIG,\,disk}$ & C$\rm _{DIG}>$0.3 spaxels    \\ 

&    [$\rm M_\odot/yr/kpc^2$] & & [$\rm erg/s/kpc^2$] & & Fraction of area   \\

\hline 
   A3128\_B\_0148   &    0.0112  &  0.75 $\pm$ 0.04 & 38.36 $\pm$ 0.04 &   0.23 $\pm$ 0.04 & 0.69  $\pm$ 0.03    \\ 
    A3266\_B\_0257   &    0.0049  &  0.57 $\pm$ 0.02 & 38.34 $\pm$ 0.03 &   0.38 $\pm$ 0.07 & 0.83  $\pm$ 0.02    \\ 
    A3376\_B\_0261   &    0.0051  &  0.80 $\pm$ 0.02 & 38.38 $\pm$ 0.02 &   0.30 $\pm$ 0.06 & 0.71  $\pm$ 0.01    \\ 
     A970\_B\_0338   &    0.0070  &  0.54 $\pm$ 0.02 & 38.46 $\pm$ 0.03 &   0.37 $\pm$ 0.05 & 0.80  $\pm$ 0.02    \\ 
           JO102   &    0.0105  &  0.71 $\pm$ 0.03 & 38.46 $\pm$ 0.04 &   0.22 $\pm$ 0.05 & 0.57  $\pm$ 0.03    \\ 
           JO123   &    0.0027  &  0.78 $\pm$ 0.08 & 38.39 $\pm$ 0.05 &   0.56 $\pm$ 0.07 & 0.87  $\pm$ 0.02    \\ 
           JO128   &    0.0022  &  0.64 $\pm$ 0.03 & 38.25 $\pm$ 0.03 &   0.51 $\pm$ 0.07 & 0.93  $\pm$ 0.01    \\ 
           JO138   &    0.0030  &  0.69 $\pm$ 0.05 & 38.15 $\pm$ 0.06 &   0.39 $\pm$ 0.10 & 0.83  $\pm$ 0.02    \\ 
           JO159   &    0.0152  &  0.50 $\pm$ 0.03 & 38.52 $\pm$ 0.06 &   0.28 $\pm$ 0.03 & 0.86  $\pm$ 0.01    \\ 
            JO17   &    0.0051  &  0.59 $\pm$ 0.01 & 38.48 $\pm$ 0.02 &   0.44 $\pm$ 0.06 & 0.86  $\pm$ 0.01    \\ 
           JO180   &    0.0027  &  0.80 $\pm$ 0.05 & 38.43 $\pm$ 0.03 &   0.50 $\pm$ 0.07 & 0.86  $\pm$ 0.02    \\ 
           JO197   &    0.0063  &  0.61 $\pm$ 0.02 & 38.41 $\pm$ 0.03 &   0.33 $\pm$ 0.05 & 0.76  $\pm$ 0.02    \\ 
           JO205   &    0.0062  &  0.71 $\pm$ 0.05 & 38.39 $\pm$ 0.05 &   0.32 $\pm$ 0.04 & 0.80  $\pm$ 0.02    \\ 
            JO41   &    0.0020  &  1.00 $\pm$ 0.10 & 38.53 $\pm$ 0.01 &   0.88 $\pm$ 0.08 & 0.98  $\pm$ 0.01    \\ 
            JO45   &    0.0017  &  0.83 $\pm$ 0.11 & 38.40 $\pm$ 0.05 &   0.63 $\pm$ 0.08 & 0.97  $\pm$ 0.01    \\ 
             JO5   &    0.0039  &  0.73 $\pm$ 0.03 & 38.31 $\pm$ 0.04 &   0.34 $\pm$ 0.05 & 0.74  $\pm$ 0.01    \\ 
            JO68   &    0.0047  &  0.50 $\pm$ 0.03 & 38.28 $\pm$ 0.05 &   0.41 $\pm$ 0.05 & 0.86  $\pm$ 0.01    \\ 
            JO73   &    0.0038  &  0.55 $\pm$ 0.03 & 38.30 $\pm$ 0.04 &   0.43 $\pm$ 0.06 & 0.93  $\pm$ 0.01    \\ 
            JO89   &    0.0015  &  0.83 $\pm$ 0.12 & 38.35 $\pm$ 0.06 &   0.87 $\pm$ 0.14 & 0.98  $\pm$ 0.01    \\ 
          P13384   &    0.0050  &  0.61 $\pm$ 0.03 & 38.24 $\pm$ 0.04 &   0.32 $\pm$ 0.05 & 0.79  $\pm$ 0.01    \\ 
          P15703   &    0.0039  &  0.81 $\pm$ 0.04 & 38.51 $\pm$ 0.03 &   0.71 $\pm$ 0.08 & 0.89  $\pm$ 0.01    \\ 
          P17945   &    0.0045  &  0.78 $\pm$ 0.03 & 38.36 $\pm$ 0.03 &   0.32 $\pm$ 0.05 & 0.74  $\pm$ 0.02    \\ 
          P20769   &    0.0037  &  0.89 $\pm$ 0.06 & 38.24 $\pm$ 0.04 &   0.39 $\pm$ 0.07 & 0.80  $\pm$ 0.02    \\ 
          P20883   &    0.0022  &  0.92 $\pm$ 0.05 & 38.33 $\pm$ 0.02 &   0.46 $\pm$ 0.08 & 0.85  $\pm$ 0.02    \\ 
          P21734   &    0.0030  &  0.71 $\pm$ 0.04 & 38.18 $\pm$ 0.05 &   0.48 $\pm$ 0.06 & 0.81  $\pm$ 0.01    \\ 
          P25500   &    0.0025  &  0.39 $\pm$ 0.04 & 37.88 $\pm$ 0.10 &   0.69 $\pm$ 0.05 & 0.97  $\pm$ 0.00    \\ 
          P42932   &    0.0086  &  0.54 $\pm$ 0.02 & 38.60 $\pm$ 0.03 &   0.40 $\pm$ 0.04 & 0.85  $\pm$ 0.01    \\ 
          P45479   &    0.0067  &  0.73 $\pm$ 0.02 & 38.59 $\pm$ 0.02 &   0.40 $\pm$ 0.06 & 0.77  $\pm$ 0.01    \\ 
          P48157   &    0.0042  &  0.56 $\pm$ 0.02 & 38.22 $\pm$ 0.03 &   0.37 $\pm$ 0.05 & 0.83  $\pm$ 0.01    \\ 
          P57486   &    0.0043  &  0.81 $\pm$ 0.03 & 38.38 $\pm$ 0.03 &   0.30 $\pm$ 0.05 & 0.73  $\pm$ 0.02    \\ 
            P648   &    0.0028  &  0.88 $\pm$ 0.05 & 38.38 $\pm$ 0.03 &   0.55 $\pm$ 0.07 & 0.83  $\pm$ 0.01    \\ 
            P669   &    0.0022  &  0.75 $\pm$ 0.08 & 38.43 $\pm$ 0.05 &   0.84 $\pm$ 0.09 & 0.98  $\pm$ 0.00    \\ 
            P954   &    0.0035  &  0.68 $\pm$ 0.05 & 38.38 $\pm$ 0.04 &   0.45 $\pm$ 0.06 & 0.87  $\pm$ 0.01    \\ 

\\
\end{tabular}  \\
\label{tab:Tab01_control}
\end{table*}

 \begin{table*}
\centering
\caption{ Stripped galaxies and their   disk $\rm \Sigma SFR$, $\rm f_0$ and  $\beta$ values, $\rm C_{DIG,\,disk}$, and the spatial fractions of spaxels that have $\rm C_{DIG}>0.3$ within the disk. }
\begin{tabular}{ccccccc}
\\
Name      & $\rm \Sigma SFR_{disk}$ & $\beta$  & $\rm log_{10}f_0$ & C$\rm _{DIG,\,disk}$ & C$\rm _{DIG}>$0.3 spaxels    \\ 

&   [$\rm M_\odot/yr/kpc^2$] & & [$\rm erg/s/kpc^2$] & & Fraction of area   \\

\hline 
       JO10   &   0.0599  &  0.70 $\pm$ 0.04 & 38.89 $\pm$ 0.05 &   0.17 $\pm$ 0.05 & 0.53  $\pm$ 0.03    \\  
           JO112   &   0.0064  &  0.64 $\pm$ 0.06 & 38.39 $\pm$ 0.06 &   0.29 $\pm$ 0.05 & 0.73  $\pm$ 0.02    \\  
           JO113   &   0.0139  &  0.51 $\pm$ 0.02 & 38.21 $\pm$ 0.04 &   0.50 $\pm$ 0.05 & 0.88  $\pm$ 0.01    \\  
            JO13   &   0.0083  &  0.51 $\pm$ 0.02 & 38.37 $\pm$ 0.03 &   0.32 $\pm$ 0.04 & 0.79  $\pm$ 0.01    \\  
           JO135   &   0.0070  &  0.68 $\pm$ 0.03 & 38.37 $\pm$ 0.03 &   0.33 $\pm$ 0.06 & 0.75  $\pm$ 0.01    \\  
           JO141   &   0.0107  &  0.54 $\pm$ 0.02 & 38.40 $\pm$ 0.03 &   0.33 $\pm$ 0.05 & 0.76  $\pm$ 0.01    \\  
           JO144   &   0.0327  &  0.52 $\pm$ 0.01 & 38.28 $\pm$ 0.03 &   0.13 $\pm$ 0.06 & 0.57  $\pm$ 0.02    \\  
           JO147   &   0.0138  &  0.53 $\pm$ 0.01 & 38.28 $\pm$ 0.02 &   0.25 $\pm$ 0.05 & 0.76  $\pm$ 0.01    \\  
           JO156   &   0.0027  &  0.66 $\pm$ 0.07 & 38.15 $\pm$ 0.08 &   0.41 $\pm$ 0.06 & 0.89  $\pm$ 0.01    \\  
           JO160   &   0.0135  &  0.48 $\pm$ 0.02 & 38.34 $\pm$ 0.05 &   0.27 $\pm$ 0.04 & 0.73  $\pm$ 0.02    \\  
           JO162   &   0.0051  &  0.57 $\pm$ 0.03 & 38.13 $\pm$ 0.05 &   0.29 $\pm$ 0.05 & 0.82  $\pm$ 0.02    \\  
           JO171   &   0.0025  &  0.38 $\pm$ 0.02 & 37.89 $\pm$ 0.06 &   0.47 $\pm$ 0.05 & 0.96  $\pm$ 0.00    \\  
           JO175   &   0.0114  &  0.49 $\pm$ 0.01 & 38.19 $\pm$ 0.02 &   0.21 $\pm$ 0.05 & 0.83  $\pm$ 0.01    \\  
           JO179   &   0.0066  &  0.66 $\pm$ 0.07 & 38.26 $\pm$ 0.09 &   0.29 $\pm$ 0.06 & 0.82  $\pm$ 0.02    \\  
           JO181   &   0.0059  &  0.93 $\pm$ 0.22 & 38.65 $\pm$ 0.11 &   0.56 $\pm$ 0.06 & 0.94  $\pm$ 0.02    \\  
           JO194   &   0.0111  &  0.30 $\pm$ 0.01 & 38.08 $\pm$ 0.04 &   0.48 $\pm$ 0.03 & 0.97  $\pm$ 0.00    \\  
           JO200   &   0.0024  &  0.71 $\pm$ 0.03 & 38.24 $\pm$ 0.03 &   0.78 $\pm$ 0.08 & 0.95  $\pm$ 0.00    \\  
           JO201   &   0.0089  &  0.71 $\pm$ 0.03 & 38.26 $\pm$ 0.03 &   0.20 $\pm$ 0.06 & 0.76  $\pm$ 0.01    \\  
           JO204   &   0.0070  &  0.60 $\pm$ 0.02 & 38.08 $\pm$ 0.03 &   0.25 $\pm$ 0.06 & 0.82  $\pm$ 0.01    \\  
           JO206   &   0.0092  &  0.47 $\pm$ 0.01 & 38.14 $\pm$ 0.03 &   0.25 $\pm$ 0.05 & 0.86  $\pm$ 0.01    \\  
            JO23   &   0.0085  &  0.64 $\pm$ 0.03 & 38.27 $\pm$ 0.05 &   0.22 $\pm$ 0.06 & 0.67  $\pm$ 0.04    \\  
            JO27   &   0.0073  &  0.30 $\pm$ 0.10 & 38.32 $\pm$ 0.26 &   0.60 $\pm$ 0.06 & 1.00  $\pm$ 0.00    \\  
            JO28   &   0.0020  &  0.80 $\pm$ 0.22 & 37.96 $\pm$ 0.19 &   0.35 $\pm$ 0.09 & 0.76  $\pm$ 0.02    \\  
            JO36   &   0.2151  &  0.61 $\pm$ 0.09 & 39.09 $\pm$ 0.12 &   0.32 $\pm$ 0.07 & 0.62  $\pm$ 0.03    \\  
            JO47   &   0.0025  &  0.58 $\pm$ 0.03 & 38.22 $\pm$ 0.04 &   0.48 $\pm$ 0.07 & 0.92  $\pm$ 0.01    \\  
            JO49   &   0.0062  &  0.57 $\pm$ 0.05 & 38.49 $\pm$ 0.06 &   0.68 $\pm$ 0.05 & 0.93  $\pm$ 0.01    \\  
            JO60   &   0.0131  &  0.48 $\pm$ 0.02 & 38.07 $\pm$ 0.05 &   0.34 $\pm$ 0.06 & 0.81  $\pm$ 0.01    \\  
            JO69   &   0.0051  &  0.61 $\pm$ 0.02 & 38.34 $\pm$ 0.03 &   0.36 $\pm$ 0.05 & 0.86  $\pm$ 0.01    \\  
            JO70   &   0.0062  &  0.66 $\pm$ 0.02 & 38.49 $\pm$ 0.02 &   0.39 $\pm$ 0.05 & 0.88  $\pm$ 0.01    \\  
            JO85   &   0.0060  &  0.51 $\pm$ 0.01 & 38.17 $\pm$ 0.02 &   0.34 $\pm$ 0.04 & 0.87  $\pm$ 0.00    \\  
            JO93   &   0.0033  &  0.68 $\pm$ 0.05 & 38.12 $\pm$ 0.07 &   0.61 $\pm$ 0.05 & 0.89  $\pm$ 0.00    \\  
            JW10   &   0.0025  &  0.52 $\pm$ 0.04 & 38.08 $\pm$ 0.06 &   0.45 $\pm$ 0.07 & 0.91  $\pm$ 0.01    \\  
           JW100   &   0.0074  &  0.55 $\pm$ 0.01 & 38.43 $\pm$ 0.02 &   0.41 $\pm$ 0.08 & 0.91  $\pm$ 0.00    \\  
           JW108   &   0.0644  &  0.52 $\pm$ 0.04 & 38.56 $\pm$ 0.08 &   0.16 $\pm$ 0.04 & 0.61  $\pm$ 0.03    \\  
           JW115   &   0.0056  &  0.94 $\pm$ 0.12 & 38.29 $\pm$ 0.09 &   0.30 $\pm$ 0.08 & 0.76  $\pm$ 0.03    \\  
            JW29   &   0.0047  &  0.74 $\pm$ 0.06 & 38.56 $\pm$ 0.05 &   0.59 $\pm$ 0.05 & 0.94  $\pm$ 0.01    \\  
            JW39   &   0.0032  &  0.62 $\pm$ 0.06 & 38.41 $\pm$ 0.07 &   0.87 $\pm$ 0.07 & 0.98  $\pm$ 0.00    \\  
            JW56   &   0.0045  &  0.96 $\pm$ 0.11 & 38.14 $\pm$ 0.07 &   0.19 $\pm$ 0.10 & 0.85  $\pm$ 0.02    \\  
\\
\end{tabular}  \\
\label{tab:Tab02_stripped}
\end{table*}

\begin{figure*}
\centering
 \includegraphics[width=0.95\textwidth]{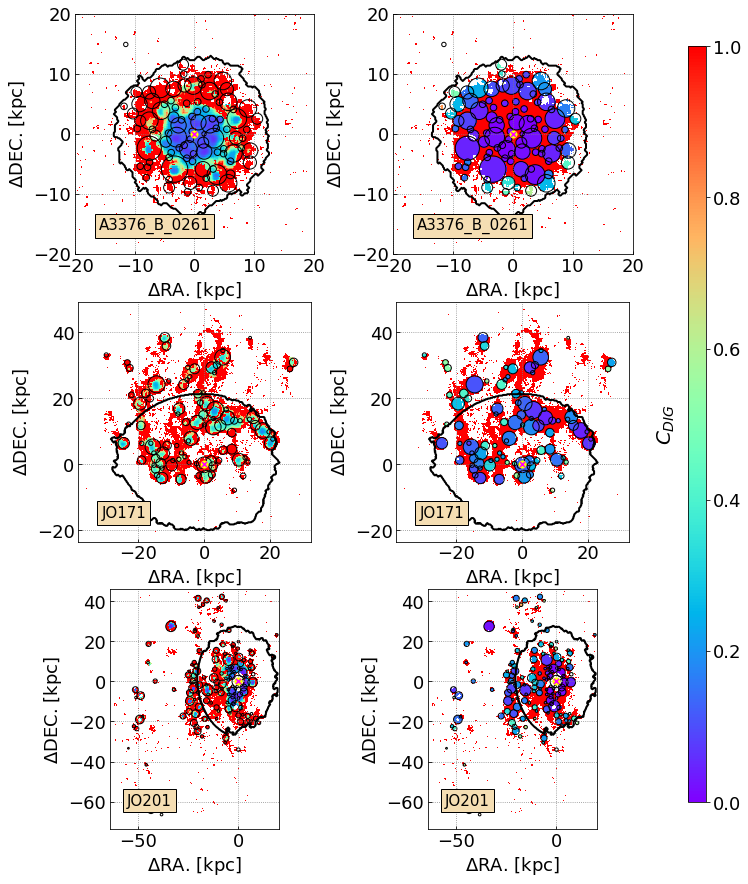}
   \caption{ \textit{Left-} Maps of emission fraction of the diffuse ionized gas ($\rm C_{DIG}$), based on the method used in this paper.We present location and sizes of  $\rm H\alpha$  clumps (estimated in \citealt{Poggianti17}) with black circles.  \textit{Right-} Maps of the fraction  of background diffuse  emission previously estimated  for the clumps. For definition of the background diffuse  emission, see Sec. \ref{subsec:Results, Cdig map}. In each map, we indicate the stellar disk (thick, black contour; \citealt{Gullieuszik20}), and  galactic center (magenta cross on yellow circle). Galaxies presented here are $A3376\_B\_0261$, JO171, and JO201.  }
    \label{fig:Cdig_clumps_maps}
\end{figure*}

\begin{figure*}
\centering
 \includegraphics[width=0.95\textwidth]{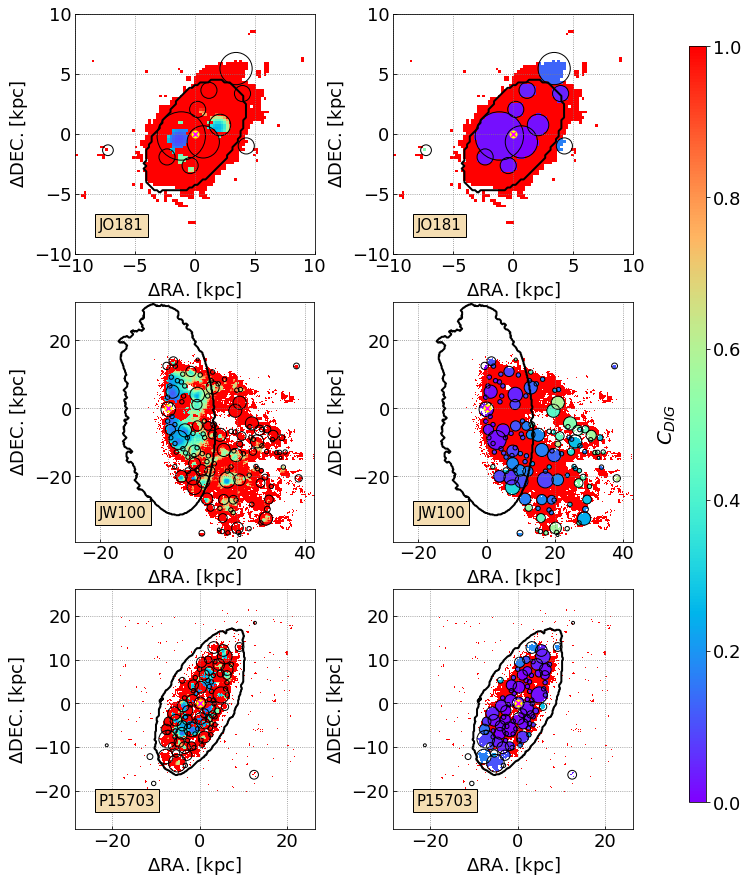}
   \caption{Same as Fig. \ref{fig:Cdig_clumps_maps} but for galaxies JO181,  JW100, and P15703.   }
    \label{fig:Cdig_clumps_maps_2}
\end{figure*}


\subsection{Potential drawbacks}
\label{subsec:Cdig drawbacks}

 Due to the kpc-scale resolution of the analyzed emission-line maps (pixel size $\approx 200$ pc, and smoothing of the data using kernels $\approx 1$ kpc  in size), we cannot resolve individual 
 H\textsc{II} regions as done in other works (\citealt{Blanc09}, \citealt{Kreckel16}, \citealt{Kaplan16}, etc.). 
Thus in each spaxel we expect contributions from both the DIG and the dense gas emission along the LOS.
This implies that spaxels dominated by dense gas emission cannot reach exactly  $\rm C_{DIG}=0$ values, but converge towards them.
However, the anti-correlation between [S\textsc{II}]/H$\alpha$ and $\rm H\alpha,corr$ seen in our data   is clear enough to    estimate $\rm C_{DIG}$ values. 

Applying this technique  at  lower spatial resolution would decrease the precision of the $\rm  C_{DIG}$ determination.
If single H\textsc{II} regions can be resolved instead, even other techniques, which use only $\rm H\alpha,corr$, or only [S\textsc{II}]/H$\alpha$ ratio, or \wha of $\rm H\alpha$ could suffice in separating DIG and dense gas dominated regions, and in estimating DIG properties.

 One assumption that we applied in Sec. \ref{subsec:estimating Cdig step by step} is that the DIG emission has the same  [S\textsc{II}]/H$\alpha$ ratio at the same metallicity. 
We note that this assumption is not strictly correct because  the DIG  shows  a wide range of [S\textsc{II}]/H$\alpha$ ratios at specific metallicities (standard deviation  $\approx0.13$; \citealt{Madsen06}). 
This increases the scatter in [S\textsc{II}]/H$\alpha$ of our data and introduces an additional uncertainty in the estimate of $\rm C_{DIG}$.

 Another drawback of our approach  is that the metallicity values were measured  using  the  PYQZ code that  assumes that the line emission (and thus line ratios)  come only from  ionization due to star-forming regions.
PYQZ does not yet have a prescription for the line ratios in the DIG that have different physical conditions compared to the \hii regions.
Thus, this may add to uncertainties in measuring the metallicities, especially in the DIG emission dominated spaxels.   

Furthermore, the metallicity radial gradient is not always symmetrical across the galaxies (A. Franchetto et al. in prep) due to inflows and morphological asymmetries in the disks and tails. 
At a given galactocentric radius, we also assume that corresponding spaxels dominated by the DIG and dense gas emission have the same metallicity which may not be correct.
The assumption in Sec. \ref{sec:estimating Cdig}  that in every galaxy we are able to always observe spaxels completely dominated by DIG emission,  may not be correct. In this case, we would be slightly overestimating the DIG fraction.
All the caveats discussed above in principle add to the scatter in the $\rm [S\textsc{II}]/H\alpha|_{Zcorr}$ values.


\section{Results} \label{sec:Results}

Following the fitting method described in Sec. \ref{subsec:estimating Cdig step by step}, using the spaxels from both the disks and tails,  we estimated single $\beta$ and $\rm f_0$ values for each galaxy. 
These values are presented in Tab. \ref{tab:Tab01_control} and \ref{tab:Tab02_stripped}, and shown in Fig. \ref{fig:beta_f0_galaxies}. 
Stripped galaxies have a statistically lower estimated $\beta$ (median 0.1  lower) compared to the control sample, but cover a similar range in $\rm f_0$. 

In what follows, we will investigate the $\rm C_{DIG}$ maps, testing how  $\rm C_{DIG}$  spaxel values (from disks of control sample and stripped galaxies) compare to  values  of $\rm \Sigma H\alpha,corr$, \sha line ratio, and $\rm W_{H\alpha}$,  in order to understand if we can use a single threshold value of those quantities to separate spaxels dominated by emission from DIG and dense gas, as typically done in other works (\citealt{Blanc09}, \citealt{Kaplan16}).
Furthermore, we  will compare integrated $\rm C_{DIG}$  with other galactic properties, for stripped and control sample galaxies.


\subsection{$\rm C_{DIG}$  maps}\label{subsec:Results, Cdig map}

The $\rm C_{DIG}$ maps for six representative galaxies are presented in the left panels of Figs. \ref{fig:Cdig_clumps_maps} and \ref{fig:Cdig_clumps_maps_2}, while  those of the rest of the sample are shown in Appendix \ref{sec:Appendix 1}.
We also over-plot the $\rm H\alpha$ clumps described in Sec. \ref{subsec: clumps}.
We remind the reader that these clumps detect peaks in $\rm H\alpha$ emission maps, and were not designated to measure fraction of emission from the DIG.
However, a single galactic value of the background diffuse  emission outside the clumps has been previously estimated as a median value of $\rm H\alpha$ surface brightness  outside the clumps (\citealt{Poggianti17}).
The fraction of  $\rm H\alpha$ flux within the clumps that come from the background diffuse $\rm H\alpha$ emission is show in the right panels of Figs.  \ref{fig:Cdig_clumps_maps} and \ref{fig:Cdig_clumps_maps_2}.   

The  $\rm C_{DIG}$ maps indicate that the  values within the clumps  are typically higher than  fractions of the background diffuse  $\rm H\alpha$ emission. 
We also notice that the $\rm H\alpha$ clumps have large radii, typically extending beyond the region that is defined as dense gas according to our method.

We now compare the fractions of the DIG emission  in the clumps  with their fractions of  background diffuse emission considering the 5202 clumps from all galaxies in the sample (control+stripped), both from the disks and the tails. 

\begin{figure}[t!]
\centering
 \includegraphics[width=0.45\textwidth]{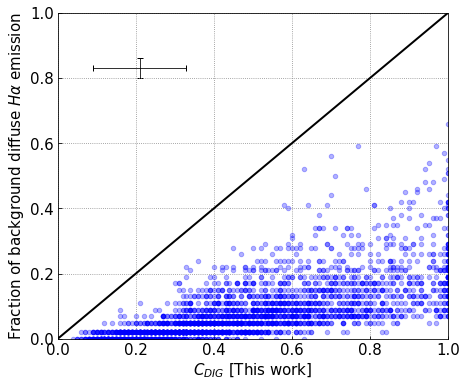}
 \caption{  Comparison  of the $\rm H\alpha$ clumps between the $\rm C_{DIG}$ values (x-axis) and  the  fractions of background diffuse emission (y-axis). For definition of the background diffuse  emission, see Sec. \ref{subsec:Results, Cdig map}.  Median uncertainties are presented by the  error bars, and $1:1$ relation is presented by the thick black line. There are 4410 clumps in total. }
    \label{fig:Fig_Clumps_Ha_SIIHa}
\end{figure}

Fig.  \ref{fig:Fig_Clumps_Ha_SIIHa} shows that in clumps  $\rm C_{DIG}$ values are higher (by a factor of $\approx5$) than the fractions of background  diffuse  emission. 
The difference  ranges from below 0.1 for the dense gas dominated clumps, to 0.8 in the DIG dominated clumps.

\begin{figure*}[ht]
\centering
 \includegraphics[width=1.\textwidth]{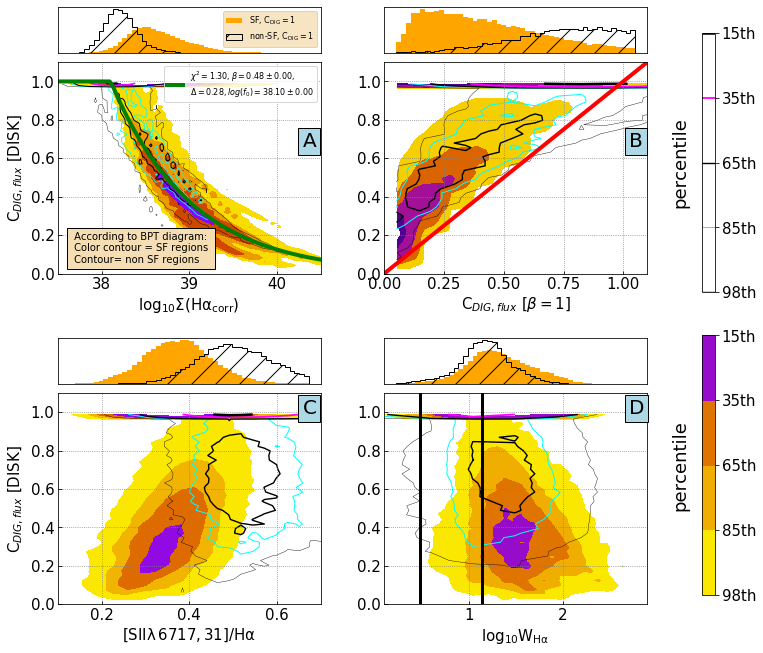}
   \caption{ Spatially resolved spaxel data (323340 spaxels in total) from the disks of all galaxies. The colored regions indicate spaxels that are designated as star-forming by the BPT diagram utilising $\rm [O\textsc{i}]/H\alpha$ line ratio, while the line-contour show non star-forming spaxels.  The contours represent the 15th, 35th, 65th, 85th and 98th percentiles.  Above each panel, we show a histogram of spaxels that have  $\rm C_{DIG}=1$. Filled (empty) histograms are for star-forming (non star-forming) spaxels.  \textit{Panel A:}  $\rm C_{DIG}$  as a function of $\rm H\alpha_{corr}$ surface brightness. The fitted  relation, following Eq. \ref{eq:Eq05}  is shown with the green line. \textit{Panel B:} Comparison between estimated $\rm C_{DIG}$ and the $\rm C_{DIG}$ calculated with  $\beta=1$  and the $\rm f_0$ that is a result of the fit in the upper-left panel. The thick red line represents 1:1 relation. \textit{Panel C:} $\rm C_{DIG}$  as a function of $\rm [S\textsc{II}]/H\alpha$ ratio. \textit{Panel D:} $\rm C_{DIG}$  as a function of equivalent width $\rm W_{H\alpha}$.  $\rm W_{H\alpha}$ equal to 3\,\AA\, and 14\,\AA\, are indicated by the thick, black lines.  }
    \label{fig:Fig_Resolved_Cdig}
\end{figure*}


\subsection{ Spaxel by spaxel comparison}
\label{subsec:Results, clumps, Ha and SIIHa}

In  Fig. \ref{fig:Fig_Resolved_Cdig} we investigate how $\rm C_{DIG}$ values from spatially resolved spaxels  behave as a function of $\rm H\alpha_{corr}$,  $\rm [S\textsc{II}]/H\alpha$, and $\rm W_{H\alpha}$ and inspect if the $\rm [O\textsc{i}]$-BPT diagram can distinguish  spaxels dominated by the dense gas  and the DIG emission. 
With the colored regions, we indicate spaxels that are designated as star-forming by the $\rm [O\textsc{i}]$-BPT diagram, while with the line-contour we show non star-forming spaxels. 
Above each panel, we show  filled (empty) histograms of star-forming (non star-forming) spaxels that have  $\rm C_{DIG}=1$.  
Here, we plot only spaxels from the disks.

In Panel A   of Fig. \ref{fig:Fig_Resolved_Cdig} we present $\rm C_{DIG}$ as a function of $\rm \Sigma H\alpha_{corr}$.
We fitted all the data following Eq. \ref{eq:Eq05}, which results in  $\rm \beta=0.48$ and $\rm f_0=1.51\,\times\,10^{38}$ $\rm erg/s/kpc^2$. 
 $\rm C_{DIG}$ has a large scatter  ($\rm \Delta C_{DIG}\approx 0.28$), therefore  using only this best fit $\rm H\alpha_{corr}$ relation can lead to errors on $\rm C_{DIG}$ estimates up to 40\%.

The contours show that both the star-forming and non-star-forming  spaxels  show a large range in $\rm \Sigma H\alpha_{corr}$ and in $\rm C_{DIG}$.
The spaxels with $\rm C_{DIG}=1$ (histograms in Fig. \ref{fig:Fig_Resolved_Cdig}) also show a large range in  $\rm \Sigma H\alpha_{corr}$  (up to 1.5 dex), with star-forming having slightly higher values. 
The star-forming and non star-forming contours overlap in almost the entire range of $\rm C_{DIG}$ and $\rm H\alpha_{corr}$.

In Panel B we consider a case where we apply $\beta=1$  and the $\rm f_0$ that is a result of the fit in the upper-left panel. 
Here we estimated $\rm C_{DIG}$ values  by applying Eq. \ref{eq:Eq05} on the $\rm \Sigma H\alpha_{corr}$  spaxel values, assuming $\beta=1$ and a single $\rm f_0$ value. 
The purpose of this plot is to test whether the assumption that the DIG surface brightness is constant across the galaxy/ies (thus $\beta=1$), and that using a single $\rm H\alpha_{corr}$ threshold value (single $\rm f_0$) would yield different results in $\rm C_{DIG}$ compared to our method. 
There is a large offset between those values and our estimated  $\rm C_{DIG}$ values, suggesting that the hypothesis of $\beta=1$ is not reliable. 

\begin{figure}[t!]
\centering
 \includegraphics[width=0.4\textwidth]{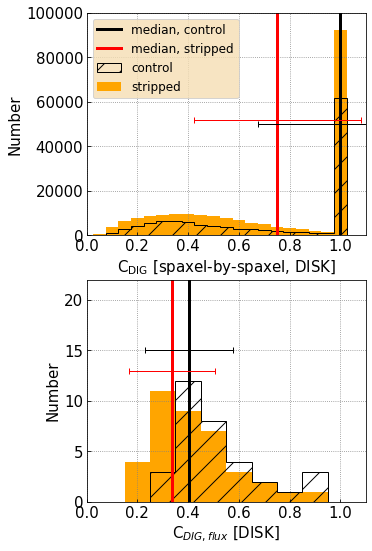}
   \caption{ Histograms of disk spaxel-by-spaxel $\rm C_{DIG}$ values (upper panel) or integrated values of  the disks ($\rm C_{DIG,\,disk}$; bottom panel) of control sample (empty histogram, with diagonal lines) and stripped galaxies (orange filled histogram). We present median values (and  corresponding $1\,\sigma$ of distributions) with red (black) lines for the control (stripped) galaxies data.  }
    \label{fig: Cdig histogram}
\end{figure}

Comparing the spaxel-by-spaxel distributions of $\rm C_{DIG}$ for the stripped galaxies and the control sample (upper panel in Fig. \ref{fig: Cdig histogram}), we find that the two distributions are similar and cover the same ranges.
It is important to note that spaxels dominated by DIG or by dense gas, respectively, do not necessarily correspond to non star-forming vs star-forming spaxels. 
In Panel C of Fig. \ref{fig:Fig_Resolved_Cdig}, it is clearly seen that star-forming spaxels preferentially have lower $\rm C_{DIG}$ than non star-forming spaxels.
However, even star-forming spaxels can have a $\rm C_{DIG}=1$, and non star-forming spaxels can reach low $\rm C_{DIG}$ values. 
Thus, the dichotomy $\rm C_{DIG}$ vs. dense gas does not correspond necessarily to the distinction non star-forming vs star-forming. 
Even regions dominated by DIG can have star formation as dominant ionization source according to the $\rm [O\textsc{i}]$-BPT diagram.
The partial overlap of the star-forming and the non-star-forming points might be related to the fact that even in regions that appear to be dominated by one ionization mechanism or another (based on the BPT diagram) there is probably a contribution from different ionization mechanisms (as discussed previously by \citealt{Poggianti19b}).

The same panel displays $\rm C_{DIG}$ values as a function of  $\rm [S\textsc{II}]/H\alpha$ ratio. 
As expected, $\rm C_{DIG}$ shows a correlation with $\rm [S\textsc{II}]/H\alpha$. 
The data shows a large scatter, thus making a single fit not viable.
Furthermore, there is a large overlapping area (both in $\rm C_{DIG}$ and $\rm [S\textsc{II}]/H\alpha$) between star-forming and non star-forming distributions.
The spaxels with $\rm C_{DIG}=1$ cover the $\rm 0.2< [S\textsc{II}]/H\alpha <0.7$ range.

In Panel D of Fig. \ref{fig:Fig_Resolved_Cdig}, we compare $\rm W_{H\alpha}$ with $\rm C_{DIG}$.
We indicate $\rm W_{H\alpha}=3$\,\AA\, and $\rm W_{H\alpha}=14$\,\AA\, values, which are  given by \citet{Lacerda18} to separate spaxels dominated by DIG emission from ones dominated by dense gas emission. 
The plot shows that most spaxels ($68\%$) have $\rm W_{H\alpha}>14$\,\AA\,, while some ($30\%$) have 3\,\AA$\rm <W_{H\alpha}<14$\,\AA.
We do not see a significant difference in distribution of  $\rm W_{H\alpha}$ for spaxels with $\rm C_{DIG}=1$ and the rest of spaxels.

The main conclusions that we draw from these results are: 1) each galaxy shows different $\beta$  and $\rm f_0$ values, 2) we cannot assume that $\beta=1$, or that the DIG $\rm H\alpha$ emission is  constant in surface brightness across galaxies,  3) we cannot use a single $\rm H\alpha_{corr}$ value, or a single  $\rm [S\textsc{II}]/H\alpha$ value, or a single $\rm W_{H\alpha}$ value to separate spaxels dominated by emission from the dense gas from the DIG, and 4) the BPT criteria identifying the source of ionisation (star formation vs no-star formation) cannot distinguish between spaxels dominated by the dense gas or the DIG emission.

\begin{figure*}[t!]
\centering
 \includegraphics[width=1.\textwidth]{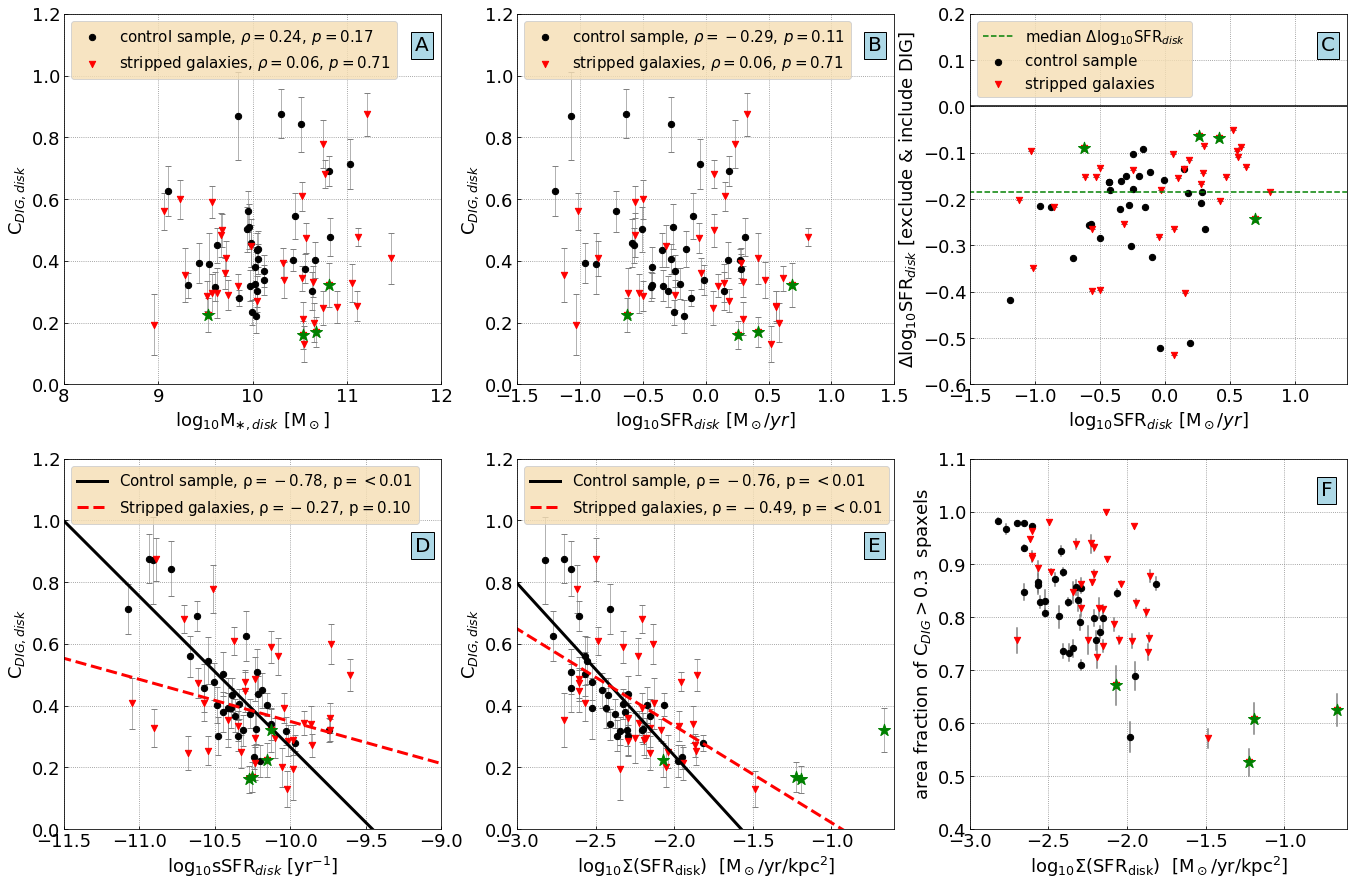}
   \caption{  Fraction of integrated $\rm H\alpha$ surface brightness coming from the DIG ($\rm C_{DIG,\,disk}$) as a function of the stellar mass (panel A), total SFR of the disk  ($\rm SFR_{disk}$; panel B), difference in $\rm SFR_{disk}$ (subtracting SFR after and SFR before removal of the DIG emission; panel C),  sSFR (panel D),  mean surface density of SFR ($\rm \Sigma SFR$; panel E). We separate and fit the data into the control sample (black circles), and stripped galaxies (red triangles), and label Pearson's correlation coefficients and corresponding p-values. In the panels E, the figure  shows area fraction of spaxels that have $\rm C_{DIG}>0.3$ (y axis), as a function of $\rm \Sigma SFR_{disk}$.    Uncertainties in values of area fraction are calculated using error estimation of a bimodal distribution. We add a mean of the area fraction of star-forming spaxels for the control (stripped) data as the black (red) vertical line.  We mark galaxies with truncated disk (JO10, JO23, JO36 and JW108) with large green   star symbols.  }
    \label{fig:Fig_Cdig_galaxies}
\end{figure*}


\subsection{Integrated DIG fraction}\label{subsec:Results, all gal}

Using the derived $\rm C_{DIG}$, and multiplying them with the $\rm \Sigma H\alpha$ maps, we derived maps of the DIG surface brightness  ($\rm \Sigma H\alpha_{DIG}$). 
We  estimated then the fraction of  the DIG emission within the disks ($\rm C_{DIG,\,disk}$) of all galaxies (control sample and stripped galaxies) as a ratio of integrated $\rm \Sigma H\alpha_{DIG}$ and integrated $\rm \Sigma H\alpha$.
Here, we only use data from the disks in order to avoid the introduction of biases,  due to the tails of the stripped galaxies.  
The galactic stellar masses ($\rm M_{\ast}$) and SFR  are presented in \citet{Vulcani18}. 
We then compute the specific Star Formation Rate (sSFR) as SFR/$M_\star$.

In Fig. \ref{fig:Fig_Cdig_galaxies}, we plot $\rm C_{DIG,\,disk}$ as a function of $\rm M_{\ast}$ (panel A), SFR of the disk ($\rm SFR_{disk}$, panel B), sSFR  (panel D), and mean  surface density  of SFR ($\rm \Sigma SFR_{disk}$) of spaxels (panel E). 
We remind the reader that we define spaxels dominated by emission from the DIG as those which have $\rm C_{DIG}>0.3$. 
 The difference in $\rm SFR_{disk}$ values (Panel C) is calculated by dividing  the SFRs of spaxels where the DIG emission was removed, and  the SFRs of spaxels with combined emission from the DIG and the dense gas.

The figure also shows the fraction of area dominated by DIG emission (area of all spaxels with H$\alpha$ emission), as a function of $\rm \Sigma SFR_{disk}$ (panel F).
Here, we define spaxels dominated by the DIG emission if their $\rm C_{DIG}$ is greater than 0.3.
This is an arbitrary choice, and changing it to higher number (for example $\rm C_{DIG}=0.6$) does not change results.
We present the values of those data in Tab. \ref{tab:Tab01_control} and \ref{tab:Tab02_stripped}.
We separate the data between the control sample and stripped galaxies, and add Pearson's correlation coefficients ($\rho$) to certain trends. 

The disk integrated $\rm C_{DIG,\,disk}$ values range  between 0.2 and 0.9, indicating that the DIG flux contributes from 20\% to 90\% in the galaxy disks.
Stripped and control sample galaxies cover a similar range in $\rm C_{DIG,\,disk}$, as also seen in bottom panel  of Fig.  \ref{fig: Cdig histogram}). The Kolmogorov-Smirnov test\footnote{Using \textit{scipy.stats.ks\_2samp} Python code.  } on those samples  results in p-value of 5.8\%, thus indicating that those two samples come from a same population.

We do not see any trend between $\rm C_{DIG,\,disk}$ and the stellar mass or SFRs (Panels A and B), as indicated by relatively low Pearson's coefficients ($|\rho|<0.3$).
As expected, subtracting the DIG emission lowers  $\rm SFR_{disk}$ by $\approx$0.2 dex (Panel C). 

On the other hand, $\rm C_{DIG,\,disk}$ seems to  anti-correlate with sSFR (Panel D), when all galaxies are considered. 
The control sample galaxies show a strong correlation ($|\rho|\approx0.78$) compared to the stripped galaxies($|\rho|\approx0.27$), with trends :

\begin{equation}
\rm C_{DIG,\,disk} = (-0.48 \pm 0.07)\cdot log_{10}sSFR + (-4.6 \pm 0.7) 
\label{eq:Eq06a}
\end{equation}

\noindent
for the control sample, and

\begin{equation}
\rm C_{DIG,\,disk} = (-0.14 \pm 0.08)\cdot log_{10}sSFR + (-1 \pm 0.8) 
\label{eq:Eq06b}
\end{equation}

\noindent
for the stripped galaxies.

There is a clear correlation also between $\rm C_{DIG,\,disk}$ and $\rm \Sigma SFR_{disk}$ (Panel E), with high Pearson's   coefficients ($\rho=-0.5$ and $\rho=-0.7$ for stripped and control galaxies). 
For the fit,  we excluded galaxies with truncated disks (JO10, JO23, JO36 and JW108; marked as green stars).
These anti-correlations (fitted using MPFIT\footnote{http://purl.com/net/mpfit}; \citealt{MPFIT09}) are as follows:

\begin{equation}
\rm C_{DIG\,disk} = (-0.56 \pm 0.09)\cdot log_{10}\Sigma SFR  + (-0.9 \pm 0.2) 
\label{eq:Eq06c}
\end{equation}

\noindent
for the control sample, and

\begin{equation}
\rm C_{DIG\,disk} = (-0.31 \pm 0.09)\cdot log_{10}\Sigma SFR  + (-0.3 \pm 0.1) 
\label{eq:Eq06d}
\end{equation}

\noindent
for the stripped galaxies.  

Generally, the data from both types of galaxies cover a similar area on the diagram, though the stripped galaxies show a larger scatter. 
The data points indicate that  the stripped galaxies show slightly higher  $\rm C_{DIG,\,disk}$ values at high $\rm \Sigma SFR_{disk}$,  but cover a similar range in $\rm C_{DIG,\,disk}$  at low $\rm \Sigma SFR_{disk}$.  

The area fraction of the DIG dominated  spaxels indicate higher values for the stripped galaxies at a given  $\rm \Sigma SFR_{disk}$, compared to the control sample (panel E in Fig. \ref{fig:Fig_Cdig_galaxies}).
That indicates that, at a given $\rm \Sigma SFR_{disk}$, the spaxels dominated by  emission from the DIG cover a larger fraction of the area in stripped galaxies, compared to control galaxies.


\section{Discussion}\label{sec: Discussion}

Here we comment on  how the $\rm C_{DIG}$ measuring  technique  used in this paper, which utilises both the $\rm H\alpha$ surface brightness and the \shar, differs from other methods used in the literature, and we discuss the benefits of our method.
We also  discuss variations and trends in integrated $\rm C_{DIG,\,disk}$ among the galaxies.


\subsection{Methods based only on H$\alpha$ or \shar, and spaxel-by-spaxel comparison}\label{subsec:Discussion, models}

Previous spatially resolved studies used methods of estimating the DIG distribution and  fractions  that assume  only a certain threshold in H$\alpha$ surface brightness,  or assume a threshold in \sha line ratio (\citealt{Oey07}, \citealt{Kreckel16}, \citealt{Zhang17}, \citealt{Poetrodjojo19}, \citealt{Kumari19}, \citealt{Brok20}). 
This is possible when a  spatial resolution (spaxels sizes $\rm <50$\,pc) enables to resolve individual HII regions and their association, such as in \citet{Kreckel16}. 
However, this is not possible at lower spatial resolutions.

Our results indicates that we cannot use a single value in H$\alpha$ surface brightness  or  \sha  to separate spaxels dominated by DIG or the dense gas as seen in Fig. \ref{fig:Fig_Resolved_Cdig}. 
For example, previously used H$\alpha$ clumps were determined using only   H$\alpha$ maps, but it resulted in fractions of the  background diffuse emission within those clumps to be  underestimated compared to the $\rm C_{DIG}$ values. 
In conclusion, combining   information from both the H$\alpha$ surface brightness and \sha line ratios, the determined $\rm C_{DIG}$ should  be measured more precisely and avoid the main drawbacks of previously used  methods. 

Although there is a spaxel-by-spaxel correlation between  $\rm C_{DIG}$ and $\rm \Sigma H\alpha,corr$ due to the relation seen in Eq. \ref{eq:Eq05}, there is a relatively large scatter caused by variations among the galaxies and their values of $\beta$ and $\rm f_0$. 

Our results also indicate that we cannot use the same $\beta$ and $\rm f_0$ value for spatially resolved data of different galaxies (upper-right panel in Fig. \ref{fig:Fig_Resolved_Cdig}).


\subsection{$H\alpha$ Equivalent width}\label{subsec:Discussion, Wha}

\begin{figure*}[ht]
\centering
 \includegraphics[width=0.95\textwidth]{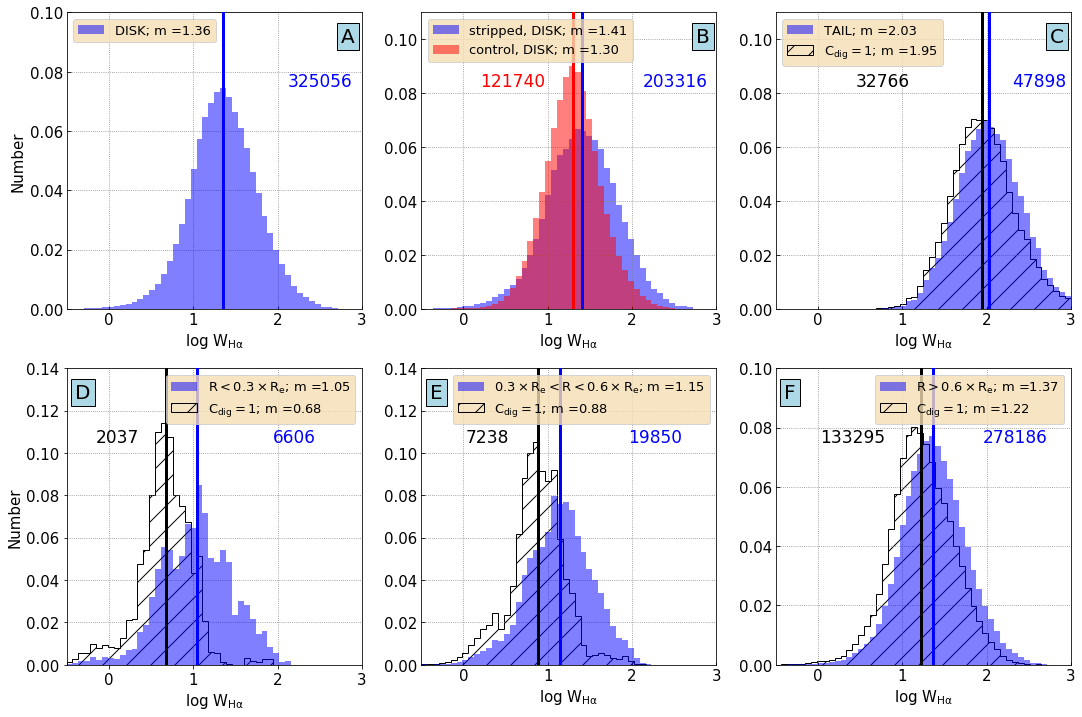}
   \caption{  Histograms of  H$\alpha$ equivalent width ($\rm W_{H\alpha}$) for all spaxels in: all disks (panel A), stripped (blue) and control (red) galaxies (panel B), tails (blue histogram in panel C). 
   In panels D, E and F we show $\rm W_{H\alpha}$ values of disk spaxels with blue histograms, separated in 3 bins of different galactocentric radii. In Panels C, D, E and F, we added hatched histograms showing  a sub-set of corresponding spaxels that have  $\rm C_{DIG}=1$. We present also median values (labeled 'm' in the legends) of corresponding histograms with lines of the same color. We show numbers of spaxels of each histogram  next to  corresponding histograms.   }
    \label{fig:Fig_WHa}
\end{figure*}

Some previous studies applied $\rm W_{H\alpha}\leqslant3$\,\AA\, (up to $\rm W_{H\alpha}\leqslant14$\,\AA\,) to disentangle spaxels dominated by emission from  DIG and spaxels dominated by emission from   the dense gas (\citealt{Lacerda18}, \citealt{Asari19}, \citealt{Asari20}, \citealt{Mingozzi20}). 
There, spaxels that probe gas outside the star-forming regions (with low H$\alpha$ equivalent width) are dominated by ionization from Hot Low-Mass Evolved Stars (HOLMES; \citealt{Stasinska2008}, \citealt{Flores-Fajardo2011}), thus having a low $\rm W_{H\alpha}$.

While the lower limit of 3\,\AA\, in $\rm W_{H\alpha}$ is physically motivated (see \citealt{CidFernandes2011}), the upper limit of $\rm W_{H\alpha} = 14$\,\AA\, is purely empirical and depends on the spatial resolution of the observations.
In order to evaluate these criteria in our sample of MUSE datacubes, in Fig. \ref{fig:Fig_WHa}, we show histograms  of $\rm W_{H\alpha}$ values in disks, stripped vs. control sample, tails and spaxels at different galactocentric distances. 
We also separate data from the disks and tails, to see if tails yield different results from the disks.
As seen on the top row of panels Fig. \ref{fig:Fig_WHa}, the peak of the $\rm W_{H\alpha}$ distribution of disk spaxels in our overall sample (panel A) is at 22.9\,\AA.
Considering only the control sample (red histogram in panel B), which is more comparable to the CALIFA and MaNGA samples of \citet{Lacerda18} and \citet{Asari19} the peak is at 19.0\,\AA, larger than the 14\,\AA\, limit found in their works, which is expected due to the higher spatial resolution of MUSE data.

In the bottom-right panel of Fig. \ref{fig:Fig_Resolved_Cdig} we already showed that using only $\rm W_{H\alpha}$ does not allow us to clearly discriminate DIG and dense gas. This can also be seen in Fig. \ref{fig:Fig_WHa}.
$\rm W_{H\alpha}$ tends to span a large range in values (3-1000\,\AA), and we  cannot distinguish DIG dominated from non-DIG dominated spaxels only using $\rm W_{H\alpha}$ distribution from the disks. 
There is no significant difference between stripped galaxies and control sample.
However, a small difference between those samples may be caused by slightly higher spatially resolved $\rm \Sigma SFR$ of the stripped galaxies compared to control sample (\citealt{Vulcani20b}), thus resulting in slightly higher  $\rm W_{H\alpha}$.

Overall,  central regions have lower $\rm W_{H\alpha}$, similar to what is observed in nearby galaxies by \citet{Lacerda18} and \citet{Asari19},  while  in the rest of the disk,  $\rm W_{H\alpha}$ increases with  galactocentric distance.
We also notice that the DIG dominated spaxels in the disks mostly show  slightly lower $\rm W_{H\alpha}$ compared to the rest of data in the same radial bin. 
However, if we only use histograms of all $\rm W_{H\alpha}$ values to distinguish DIG dominated spaxels from the rest, we are able to do it only in the central regions (Panel D, and partly Panel E). 
In the centers, the DIG dominated spaxels have $\approx0.3$ dex lower $\rm W_{H\alpha}$ compared to the rest of data, a trend  similarly seen by \citet{Lacerda18}.

The method from \citet{Lacerda18} is  not suited   for the debris tails of the stripped galaxies. 
In the tails, $\rm W_{H\alpha}$ is $\approx0.6$ dex higher than in the disks, and there is no difference between DIG dominated and non-DIG dominated spaxels.
The tails of the gas-stripped galaxies do  not have stars  older than a few times $10^8$ yr, and its light is mostly dominated by a bright gas emission and younger stellar populations ($\rm<2-300$ Gyr, traced by ultra-violet emission; \citealt{Bellhouse17}, \citealt{George18}). 
This would drastically increase $\rm W_{H\alpha}$. 
Thus, \wha would be an  unreliable  estimator of the DIG and its fraction in the GASP galaxies, especially outside their disks.

To conclude, the variation in  $\rm W_{H\alpha}$ is not caused only by an increase in $\rm C_{DIG}$, but also potentially  by other sources. Therefore, we cannot prescribe a simple method of measuring $\rm C_{DIG}$ from $\rm W_{H\alpha}$, especially in the tails of stripped galaxies. 
 Moreover, the kpc-scale   spatial resolution of our data may also affect these measurements, due to combining regions with both low and high  $\rm W_{H\alpha}$ in the LOS.


\subsection{Integrated DIG correlations}\label{subsec:Discussion, cdig galaxies}

Our results of integrated disk data show that the DIG contributes to a large fraction of the total  H$\alpha$ flux ($\rm C_{DIG,\,disk}$) in galactic disks (between 20\% and 90\%;  Fig. \ref{fig: Cdig histogram} and \ref{fig:Fig_Cdig_galaxies}).
This is similar to the range found in many other studies, where the DIG fraction was found to contribute between 30\% and 80\% (with mean values around 50\%-60\%; \citealt{Hoopes03}, \citealt{Oey07}, \citealt{Sanders17}, \citealt{Poetrodjojo19}, \citealt{Bruna20}).

In this work, we present for the first time a comparison between the stripped galaxies, and the control sample that are non-stripped galaxies. 
Also, for the first time, we compare integrated $\rm C_{DIG,\,disk}$ with global galactic values such as  stellar mass, SFR, and $\rm sSFR_{disk}$. 

On one hand, we do not see trends between the DIG fraction and the galactic stellar mass, or SFR.
\citet{Vulcani18} found a   0.2 dex difference in integrated SFR-M$_\ast$ relation between stripped and control samples of galaxies. 
This difference in SFR-M$_\ast$ relation would still exist  if we subtract the DIG emission because the integrated SFR values would similarly decrease for both the stripped and control samples (Panel C in Fig. \ref{fig:Fig_Cdig_galaxies}).

On the other hand, we detect anti-correlations  with  $\rm sSFR_{disk}$ and $\rm \Sigma SFR_{disk}$.  
The control sample shows stronger  anti-correlation compared to the stripped galaxies. 
These anti-correlations potentially enables to  robustly, but not precisely,  measure  fraction of the   DIG emission  in galactic disks by simply using  $\rm sSFR_{disk}$ or $\rm \Sigma SFR$ values.

\citet{Oey07} and \citet{Sanders17} found a similar anti-correlation between the fraction of the DIG and $\rm \Sigma H\alpha$. 
\citet{Oey07} defined borders of bright H$\alpha$ regions as borders of dense gas regions and a background emission (constant in $\rm \Sigma H\alpha$) as a DIG emission, while \citet{Sanders17} designated spaxels within the first  10th percentile of $\rm \Sigma H\alpha$  distribution as DIG dominated spaxels. 
Furthermore, to explain spaxel-by-spaxel variation in dust attenuation and its effects on measuring Balmer line emission,    \citet{ValeAsari20} speculated that galaxies with lower $\rm sSFR_{disk}$  should have larger fraction of the DIG, the trend that we see for the galaxies in our sample.
Since $\rm sSFR_{disk}$  can be used as an age indicator of stellar population, this result may indicate a correlation between the DIG fraction and the amount of old stellar population such as HOLMES (ages of above $10^8$ yr; \citealt{Flores11}), and could explain slightly stronger correlation in the case of the control sample. 
In other words, ionization from the older stellar population may contribute to ionization of the DIG. 

At a given $\rm \Sigma SFR_{disk}$, the stripped galaxies have higher fraction of area covered by DIG dominated spaxels than the control sample.
That may indicate two possible conclusions: 1)  the DIG is brighter in surface brightness (higher $\rm \Sigma SFR_{disk}$) in the stripped galaxies at a given area fraction, or 2) the dense gas areas are smaller in size for the stripped galaxies compared to the control sample.

When we compare $\beta$ and $\rm f_0$ values for individual galaxies (Fig. \ref{fig:beta_f0_galaxies}), we notice that the stripped galaxies have median  of  $\beta \approx0.1$ lower   compared to the control sample, despite that they cover a large range ($\beta$ from 0.2  to 1). 
This may indicate that the DIG emission in galaxies (and even more in stripped galaxies) is not a constant in surface brightness across the disks. 
On what does its emission depend (on the $\rm H\textsc{II}$ regions, environment, and/or other sources) will be investigated in future papers.


\section{Summary}\label{sec: Summary}

The diffuse ionised gas (DIG) is an important component of the ISM that is affected by physical processes across the galaxies and their evolution. 
Measuring its distribution and fraction in surface brightness allows us to properly study its source of ionization and  star formation in galaxies.
Subtracting the DIG emission from the observed galactic images would remove biases from observations of the star formation  and gas-phase metallicities. 

In this paper, we measure for the first time the DIG emission in gas-stripped galaxies at different stages of gas-stripping, and compare them to normal galaxies.  
We utilise the IFU (MUSE spectrograph)  observations of galaxies from the multi-wavelength project GASP, and study 71 galaxies. We used emission line maps to estimate the fraction of emission from the DIG ($\rm C_{DIG}$) using both the H$\alpha$ and \sha line ratios (Sec. \ref{sec:estimating Cdig}).
Unlike in previous works in the literature,  we corrected the \sha ratio for metallicity gradients because our observations cover the entire galaxy disk.  

Our analysis indicates that we cannot use a single H$\alpha$ threshold values or a single \sha value to separate spaxels dominated by emission from the dense gas or the DIG (Fig. \ref{fig:Fig_Resolved_Cdig}). 
Furthermore, assuming that the DIG has a constant background emission across galaxies yields lower  $\rm C_{DIG}$ values compared to the values derived with our method. 

Also the equivalent width of H$\alpha$ cannot be used as estimator of  \cdig across disks of the entire galaxies (Fig. \ref{fig:Fig_WHa}).
At larger distances, $\rm W_{H\alpha}$ values are very high ($\rm W_{H\alpha}>14$\,\AA)  even for the DIG dominated spaxels.  
In the debris tails of the stripped galaxies, almost all spaxels show $\rm W_{H\alpha}>14$\,\AA, which we ascribe to the fact that the tails lack the older stellar population.

We compared for the first time  the DIG fractions between the stripped and non-stripped galaxies (Fig. \ref{fig: Cdig histogram} and  \ref{fig:Fig_Cdig_galaxies}). 
In both samples, the fraction of emission from the DIG  in the galactic disks ($\rm C_{DIG,\,disk}$) show a range between 20\% and 90\% of the total integrated flux. 
The $\rm C_{DIG,\,disk}$ does not  correlate with either the galactic stellar mass or the SFR.
The relative difference in SFR-M$_\ast$ relation between stripped and control samples  would not change if we subtract the DIG emission because the change in SFR values is similar in both samples.

The $\rm C_{DIG,\,disk}$ does show anti-correlations with the sSFR and the $\rm \Sigma SFR$. 
This potentially enables to  robustly measure the fraction of the  DIG emission in galactic disks by simply using $\rm sSFR_{disk}$ or $\rm \Sigma SFR$ values.  
The anti-correlation with the sSFR may be caused by a correlation between the DIG emission and  old stellar populations (older than $10^8$ yr) such as HOLMES, thus indicating that its ionization contributes to the ionization of the DIG. 
Moreover, at a given $\rm \Sigma SFR$, the DIG dominated spaxels cover a higher percentage of area in the stripped galaxies compared to the control sample.

In the following papers, we will use these estimated \cdig maps to separate galactic areas dominated by DIG and dense gas, in order to investigate the physical processes  giving rise to the DIG.
Furthermore, we will investigate in detail the emission in the tails and contrast it with the disks, to study its physical properties for different stages of stripping.

\acknowledgments

The authors wish to kindly thank Giovanni Fasano, who estimated positions and radii of H$\alpha$ clumps for GASP project. 
Based on observations collected at the European Organization for Astronomical Research in the Southern Hemisphere under ESO programme 196.B-0578. This project has received funding from the European Research Council (ERC) under the European Union's Horizon 2020 research and innovation programme (grant agreement No. 833824).
We acknowledge financial contribution from the contract ASI-INAF n.2017-14-H.0, from the grant PRIN MIUR 2017 n.20173ML3WW\_001 (PI Cimatti) and from the INAF main-stream funding programme (PI Vulcani). 
J.F. acknowledges financial support from the UNAM-DGAPA-PAPIIT IN111620 grant, Mexico.
This work  made use of Astropy, a community-developed core Python package for Astronomy  (\citealt{Astropy13}), and MPFIT (\citealt{MPFIT09}).

\facility{ VLT:Yepun (MUSE)}

\software{Astropy \citep{Astropy13}, MPFIT \citep{MPFIT09} }

\clearpage

\appendix

 \section{ $\rm C_{DIG}$ maps of all galaxies}
 \label{sec:Appendix 1}

\begin{figure*}[h!]
\centering
 \includegraphics[width=1.\textwidth]{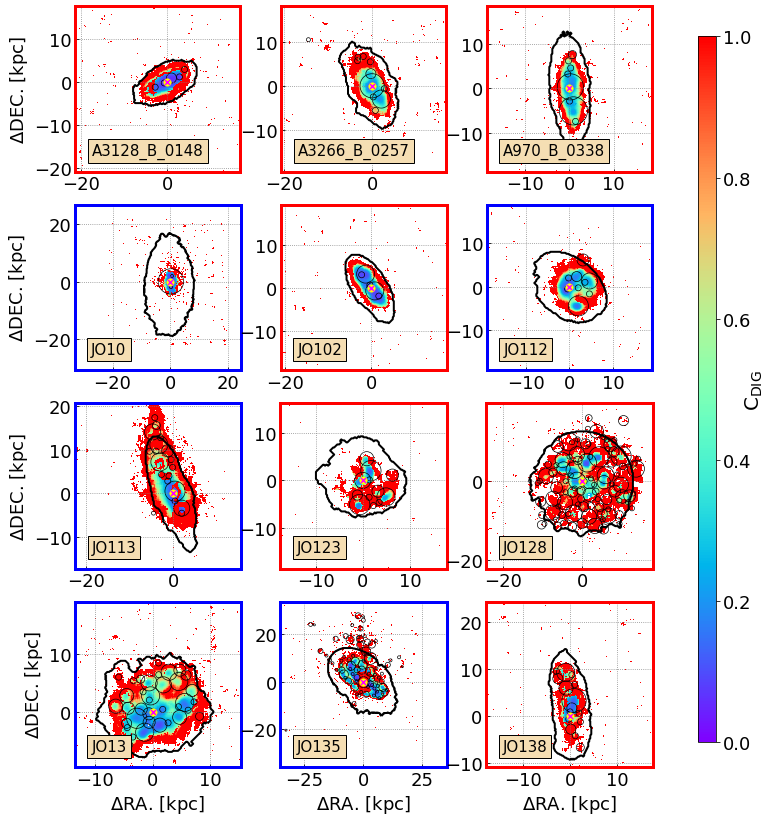}
   \caption{ Same as Fig. \ref{fig:Cdig_clumps_maps}, but for all galaxies. Here, we put panels with control sample with red edges,  and stripped galaxies  with  blue edges, respectively.     }
    \label{fig:App Cdig_maps v1}
\end{figure*}

\begin{figure*}[h!]
\centering
 \includegraphics[width=1.\textwidth]{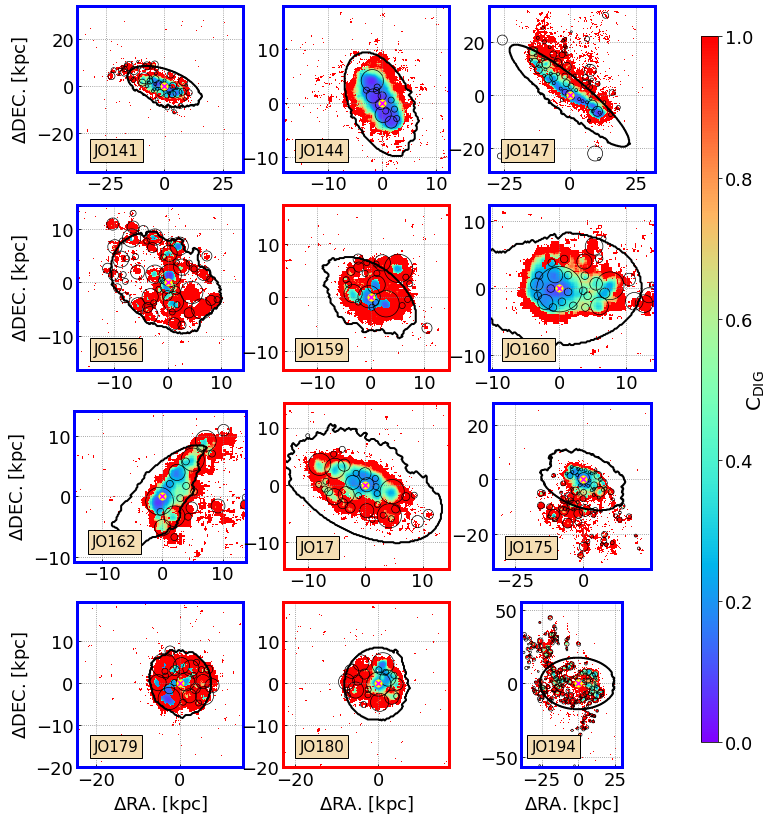}
   \caption{ Same as Fig. \ref{fig:Cdig_clumps_maps}, but for all galaxies.   }
    \label{fig:App Cdig_maps v2}
\end{figure*}

\begin{figure*}[h!]
\centering
 \includegraphics[width=1.\textwidth]{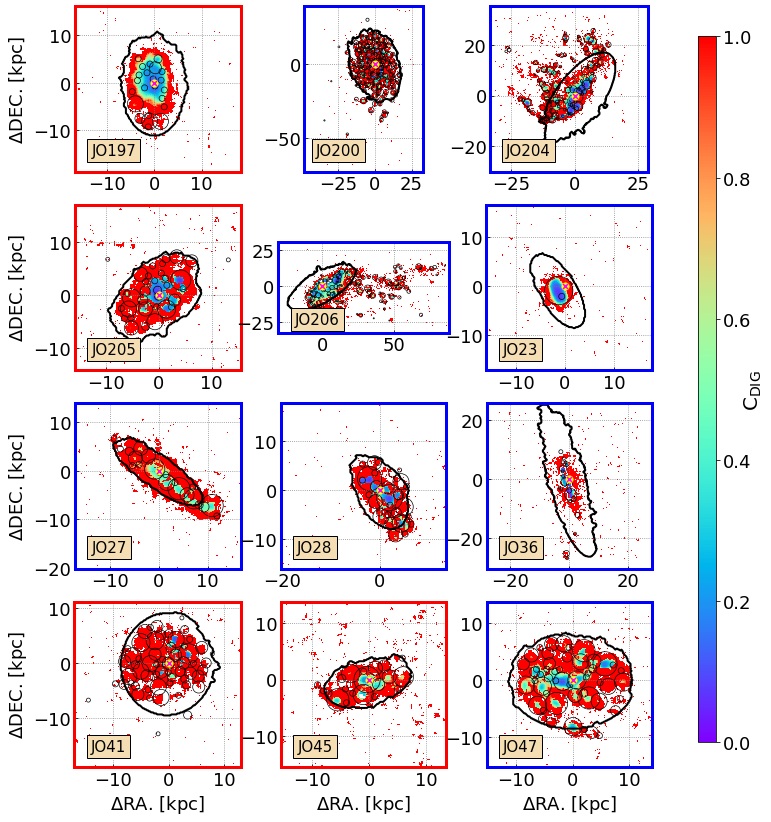}
   \caption{ Same as Fig. \ref{fig:Cdig_clumps_maps}, but for all galaxies.   }
    \label{fig:App Cdig_maps v3}
\end{figure*}

\begin{figure*}[h!]
\centering
 \includegraphics[width=1.\textwidth]{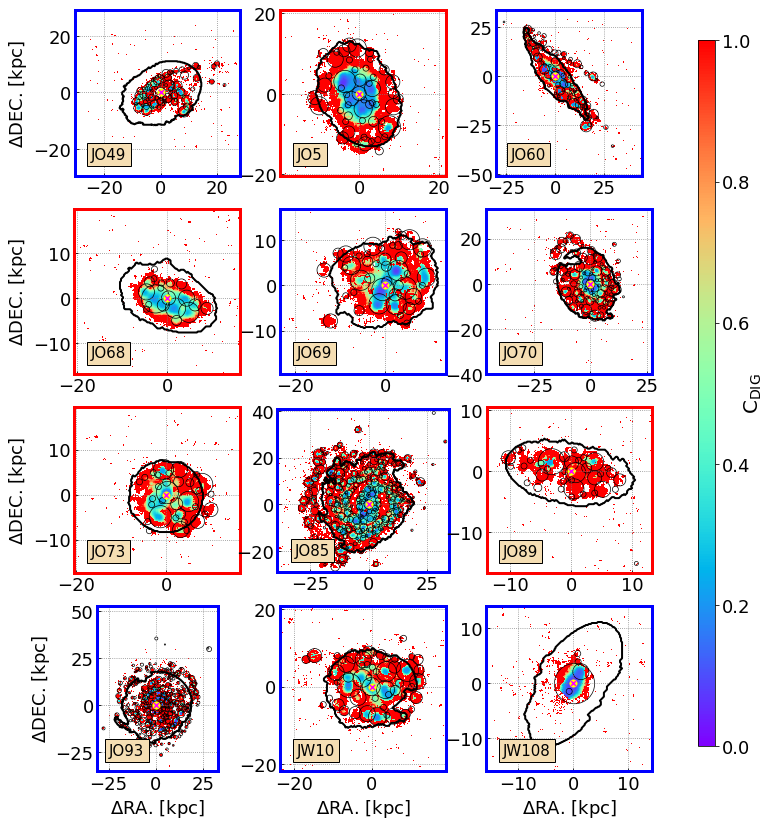}
   \caption{ Same as Fig. \ref{fig:Cdig_clumps_maps}, but for all galaxies.   }
    \label{fig:App Cdig_maps v4}
\end{figure*}

\begin{figure*}[h!]
\centering
 \includegraphics[width=1.\textwidth]{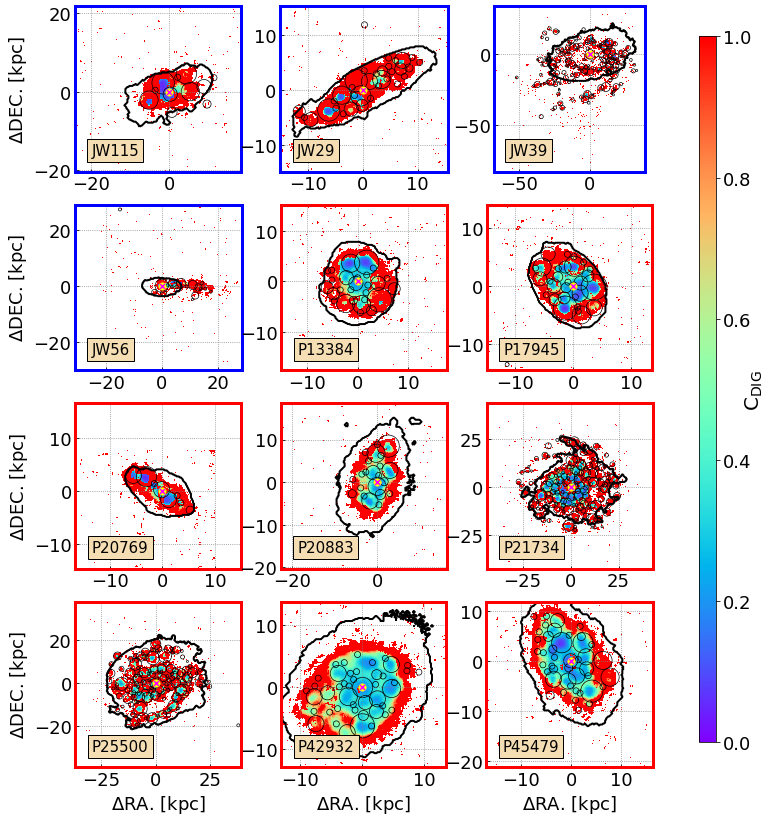}
   \caption{ Same as Fig. \ref{fig:Cdig_clumps_maps}, but for all galaxies.   }
    \label{fig:App Cdig_maps v5}
\end{figure*}

\begin{figure*}[h!]
\centering
 \includegraphics[width=1.\textwidth]{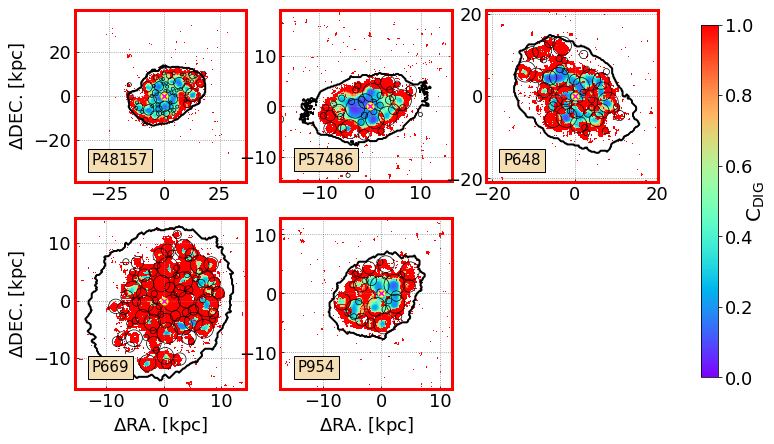}
   \caption{ Same as Fig. \ref{fig:Cdig_clumps_maps}, but for all galaxies.  }
    \label{fig:App Cdig_maps v6}
\end{figure*}

Here we present $\rm C_{DIG}$ maps of all the other galaxies in the sample, not shown in Fig. \ref{fig:Cdig_clumps_maps} (Fig. \ref{fig:App Cdig_maps v1}, \ref{fig:App Cdig_maps v2}, \ref{fig:App Cdig_maps v3}, \ref{fig:App Cdig_maps v4}, \ref{fig:App Cdig_maps v5}, and \ref{fig:App Cdig_maps v6}).
Here, we put panels with control sample and stripped galaxies with red or blue edges, respectively.

\bibliographystyle{aasjournal}
\bibliography{NT_DIG_paper}

\end{document}